# Planet formation and the evolution of the Solar System

M M WOOLFSON
University of York, Heslington, York YO10 5DD, UK[1]

**Abstract**

The Capture Theory gives planet production through a tidal interaction between a condensed star and a diffuse protostar within a dense embedded cluster. Initial extensive and highly eccentric planetary orbits round-off and decay in a circumstellar disk of material captured from the protostar. Collapsing protoplanets leave behind a circumplanetary disk within which satellites form by an accretion process. Many properties of exoplanets – orbits very close to and very far from stars, highly eccentric orbits, planets around binary stars, the proportion of stars with planets and spin-orbit misalignments are straightforwardly explained in terms of this model. It is proposed that the initial Solar System contained six major planets, the existing four plus Bellona of mass 2.5 $M_J$ and Enyo of mass 1.9 $M_J$, where $M_J$ is the mass of Jupiter. The products of a collision between the two additional planets explain many features of the Solar System – the larger terrestrial planets, Mars and Mercury and their characteristics, the Earth-Moon relationship and the Moon's surface features, the formation of asteroids, comets and dwarf planets, the formation of the Kuiper Belt and Oort Cloud, the relationship between Neptune, Pluto and Triton, the characteristics of ice-giants and isotopic anomalies in meteorites. All the mechanisms involved in these processes are well understood and occur in other astronomical contexts.

## 1 Introduction

When the Solar System was the only planetary system known, and there was no evidence that planets existed around other stars, theories of the origin of the Solar System were not constrained by the need that a theory should make planetary systems commonplace. For more than 50 years two theories have been developed, initially concerned with solar-system formation but which now, with the knowledge of the existence of several thousand exoplanets, have become general theories of planet formation. The first of these, the Nebula Theory (NT), which is the present standard and generally-accepted theory, is a monistic theory that proposes that both a star and its planets derived from a single gaseous nebula. The second, less well known, is the Capture Theory (CT), a dualistic theory that proposes that the star and planets were derived from separate sources of material. The temporal development of various aspects of the CT has been somewhat haphazard, influenced by the constraints and information from new observations as they occurred, so the purpose of this review is to describe in a logical sequence how each step in the formation of planets has a causal relationship to what preceded it. This causally-related chain of events leads further to a scenario that explains many present features of the Solar System in some detail. It is not the purpose of this review to make judgemental comparisons between the CT and NT but the difference in the approaches of the two models will be described in appropriate contexts.

---
[1] email: mmw1@york.ac.uk

## 2 Star formation

A supernova both compresses neighbouring interstellar medium (ISM) by a shock wave and injects material into it, some of which condenses into mainly-submicron dust. Both these effects give cooling of the affected region of the ISM, due to radiative cooling of the dust (Hayashi, 1966) and cooling via the excitation of molecules, atoms and ions by electron collisions (Seaton, 1955). The cooling rate increases with increasing density and reduces with reducing temperature. Cooling reduces the local pressure and the consequent influx of ISM material further increases the density. Eventually a high-density, low-temperature region is produced in pressure equilibrium with the external ISM, which is at a lower density and higher temperature. This process of producing a dense cool cloud (DCC) has been modelled using smoothed-particle hydrodynamics (SPH) by Golanski and Woolfson (2001).

If the mass of the cloud exceeds the Jeans critical mass (Jeans, 1902) then it will begin a free-fall collapse. Some of the gravitational energy released by the collapsing cloud is transformed into turbulence, the evidence for which comes from Doppler-shift measurements of maser emissions from star-forming regions, indicating turbulent motions with speeds up to 20 km s$^{-1}$ (Cook, 1977). Colliding turbulent gas streams generate high-density, high-temperature regions but, since cooling is a much faster process than re-expansion, a high-density cool region is produced that may, under suitable conditions, collapse to form a star (Woolfson, 1979). If the angular momentum in the compressed region is sufficiently high then it may bifurcate to form a binary pair where the majority of the angular momentum is taken up in the orbits of the constituent stars rather than in stellar spin (Woolfson, 2011a).

The Sun, and main-sequence stars of similar and lower mass, spin slowly with the equatorial speed of the Sun being just 2 km s$^{-1}$. However, original equatorial speeds may have been much higher, the only constraint being that the star should be rotationally stable. During the T-Tauri stage of a star's development very strong stellar winds occur, the escaping ionized material of which is linked to stellar magnetic field lines and it is carried out to a distance of several stellar radii at a constant angular speed before it decouples. This gain of angular momentum by the stellar wind leads to a reduction of angular momentum of the star and the majority of the original stellar angular momentum may be removed in this way (Cole and Woolfson, 2013).

The progress of star formation in a galactic cluster was investigated by Williams and Cremin (1969) by observing Young Stellar Objects (YSOs) in four very young clusters. From the position of YSOs on a Hertzsprung-Russell diagram it is possible to deduce both their masses and their ages from the time when they first became identifiable protostars. The result of this investigation is shown for the young cluster NGC2264 in Figure 1. This shows that:

(i) The first stars are produced about 8 $\times 10^6$ years ago. The earlier group are thought to be aberrant
(ii) The first stars produced have an average mass somewhat greater than 1 $M_\odot$
iii) There are two streams of development, one with reducing mass with time and the other, starting about 5 million years ago, with increasing mass with time.
(iv) The rate of formation of stars increases with time

The stream of increasing mass with time can be related to the model of Bonnell, Bate and Zinnecker (2005) for the formation of massive stars by the aggregation of smaller mass stars or protostars in a dense stellar environment.

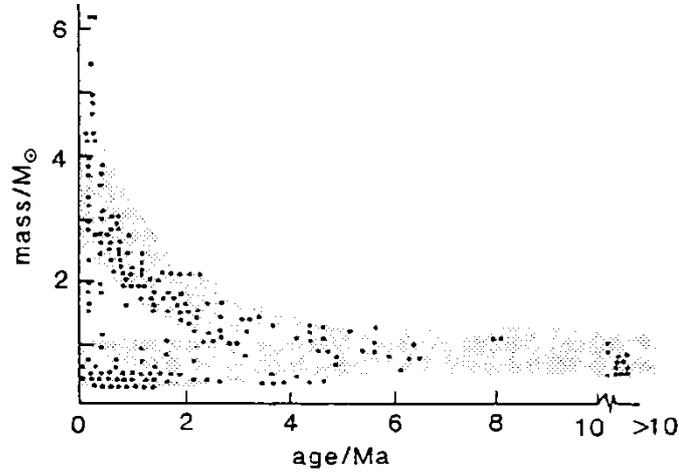

Figure 1   The masses of stars produced in a young stellar cluster as a function of time. The origin represents 'now' (After Williams and Cremin, 1969)

A typical newly-formed protostar could have a radius of 2000 au, density $10^{-14}$ kg m$^{-3}$ and temperature 20 K, corresponding to a mass of somewhat over 0.5 $M_\odot$. The free-fall time for such a body is $t_{ff}$ = 21,000 years and it will remain an extended object for most of that period; after a time 0.8 $t_{ff}$ (~ 17,000 years) its radius will have fallen to 1,000 au.

A forming cluster of stars, immersed in the gas of the cloud, is said to be in an *embedded state*. As stars form, the cloud, together with the contained stars, continues to collapse under self-gravity so that the stellar number density (SND) steadily increases. Radiation from the forming stars slowly expels gas from the cloud but, when the most massive stars become supernovae after a few million years, the rate of gas expulsion greatly increases. Released from the gravitational influence of the gas the cluster of stars begins to expand.  In 90% of cases it will expand indefinitely to give field stars and field binary systems.  In the other 10% of cases a galactic stellar cluster, typically containing a few hundred stars, is produced in quasi-equilibrium; over the course of $10^8$ or more years it will slowly evaporate ending with a small stable stellar system.

The maximum SND reached in the dense embedded state of a cluster can be extremely high; the core of the Trapezium Cluster within the Orion Nebula is estimated to have an SND several times $10^4$ pc$^{-3}$. A simulation of the evolution of a star-forming cloud by Bonnell, Bate and Vine (2003), using high-definition SPH, showed that in the last stages of the collapse, of duration about 5 million years, the cloud broke up into fragments each containing tens of stars within which the SND was up to $2 \times 10^5$ pc$^{-3}$, although the whole-cloud average SND was two orders of magnitude less. Subsequently the fragments expand and combine to form larger fragments but these larger fragments move closer together so that although the peak SND within fragments decreases the whole-cloud average SND increases. At the end of the simulation there were 400 stars in five fragments with the maximum fragment SND somewhat greater than $2 \times 10^4$ pc$^{-3}$ and whole-cloud SND peaking at $2 \times 10^4$ pc$^{-3}$.

The average speed of stars in a dense embedded cluster is about 1 km s$^{-1}$ (Gaidos, 1995) so in 17,000 years, during which time it is an extended object of radius between 1,000 and 2,000 au, a protostar can travel more than 3,000 au. For an SND of $2 \times 10^4$ pc$^{-3}$ the average distance between stars is just over 8,000 au. From the dimensions of protostars, the distances they travel as extended objects and the SNDs that can occur it is clear that close approaches of extended protostars and condensed stars can take place. What we now consider is a tidal interaction between an extended protostar and a compact YSO or main-sequence star. The frequency of such interactions is considered quantitatively in Section 8.

## 3  Capture-theory simulation

Some mechanisms involved in the CT were also an intrinsic part of the tidal theory of solar-system formation proposed by Jeans (1917). This theory envisaged that a massive star, passing close to the Sun, drew up a huge solar tide that emerged from the Sun in the form of a filament. The filament was gravitationally unstable and broke up into a string of blobs that eventually condensed to form planets. The blobs were attracted by the retreating massive star and were left in orbit around the Sun. This was the standard theory of the formation of the Solar System for two decades. Although similar mechanisms occur in the CT it is in a completely different context.

Figure 2 shows a CT simulation using SPH with radiation transfer (Oxley and Woolfson, 2003; 2004)

*Characteristics of the star*
Mass of the star, $M_* = 2 \times 10^{30}$ kg $\approx M_\odot$
Luminosity of the star, $L_* = 4 \times 10^{26}$ W $\approx L_\odot$

*Characteristics of the protostar*
Mass of the protostar, $M_P = 7 \times 10^{29}$ kg $\approx 0.35\, M_\odot$
Initial radius of the protostar, $R_P = 800$ au
Initial temperature of the protostar, $T_P = 20$ K
Mean molecular mass of protostar material, $\mu = 4 \times 10^{-27}$ kg

*Characteristics of the protostar orbit*
Initial distance of centre of protostar from star, $D = 1{,}600$ au
Closest approach of protostar orbit to star, $q = 600$ au
Eccentricity of the orbit, $e = 0.95$

At this closest-approach distance, virtually the whole protostar is distorted into a dense filament. As shown theoretically by Jeans, the filament is gravitationally unstable and breaks up into a string of blobs. These blobs have greater than the Jeans critical mass and collapse to become condensed objects (Figure 3). Five blobs, with masses 4.7 $M_J$, 7.0 $M_J$, 4.8 $M_J$, 6.6 $M_J$ and 20.5 $M_J$ ($M_J$ is mass of Jupiter}, are captured into orbit around the star – hence the name of the theory. Some blobs are not captured and are released into interstellar space as *free-floating planets* (Lucas and Roche, 2000; Sumi et al., 2011).

In the star-forming cloud there are collisions of turbulent gas streams that may not progress to form protostars but, nevertheless, if the collision takes place in the vicinity of a condensed star they can give rise to planet formation. In Figure 4 each stream has mass of 0.5 $M_\odot$, density $4 \times 10^{-15}$ kg m$^{-3}$ and speed 1 km s$^{-1}$. The condensations A, B and C, with masses 1.0 $M_J$, 1.6$M_J$ and 0.75 $M_J$ respectively, are captured.

The capture-theory mechanism is very robust. With appropriate densities of the diffuse body – the protostar in Figure 2 – and size of orbit it gives planet formation at scales from one tenth to ten times the scale used in Figures 2 and 4.

While the formation of planets within the mass range found for exoplanets has been illustrated here, the initial orbits of the protoplanets do not correspond to what is found for exoplanets or solar-system planets. For the eight protoplanets captured in the Figures 2 and 4 simulations the semi-major axes, $a$, and eccentricities, $e$, are as follows:

($a$, $e$) = (1,247 au, 0.835)  ($a$, $e$) = (1,885 au, 07725)  ($a$, $e$) = (1,509 au, 0.765)
($a$, $e$) = (1,325 au, 0.736)  ($a$, $e$) = (2,686 au, 0.902)  ($a$, $e$) = (4,867 au, 0.768)
($a$, $e$) = (1,703 au, 0.381)  ($a$, $e$) = (1,736 au, 0.818)

These orbits are more extensive than any found for exoplanets and their eccentricities are large although, as we shall see, some exoplanet orbits do have large eccentricities

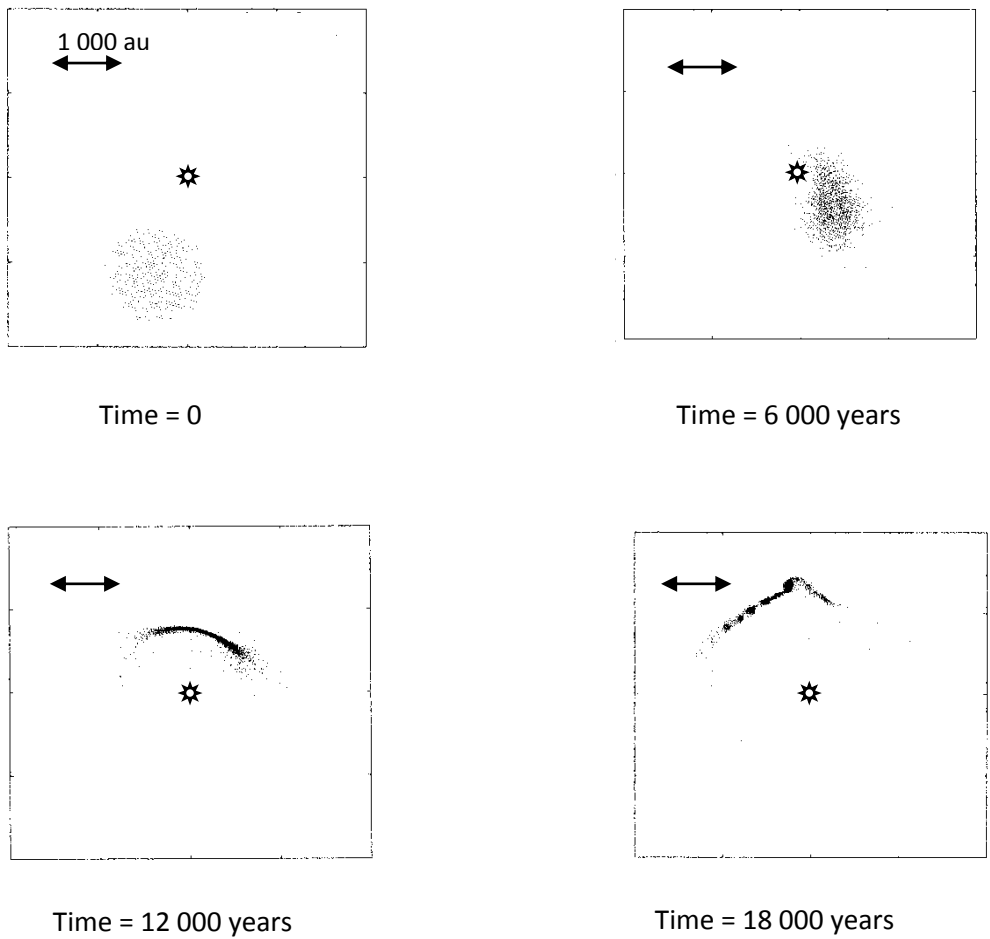

Figure 2  An SPH simulation, with radiation transfer, of the interaction of a star and a protostar (Oxley and Woolfson, 2004)

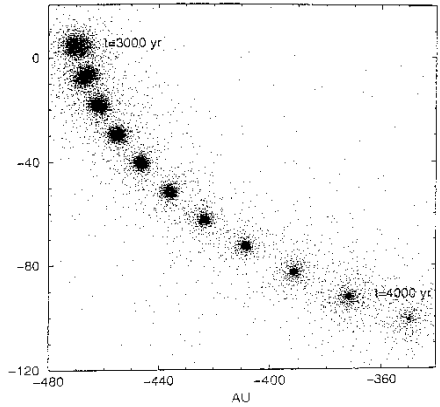

Figure 3 The collapse of a protoplanet at 100 year intervals (Oxley and Woolfson, 2004)

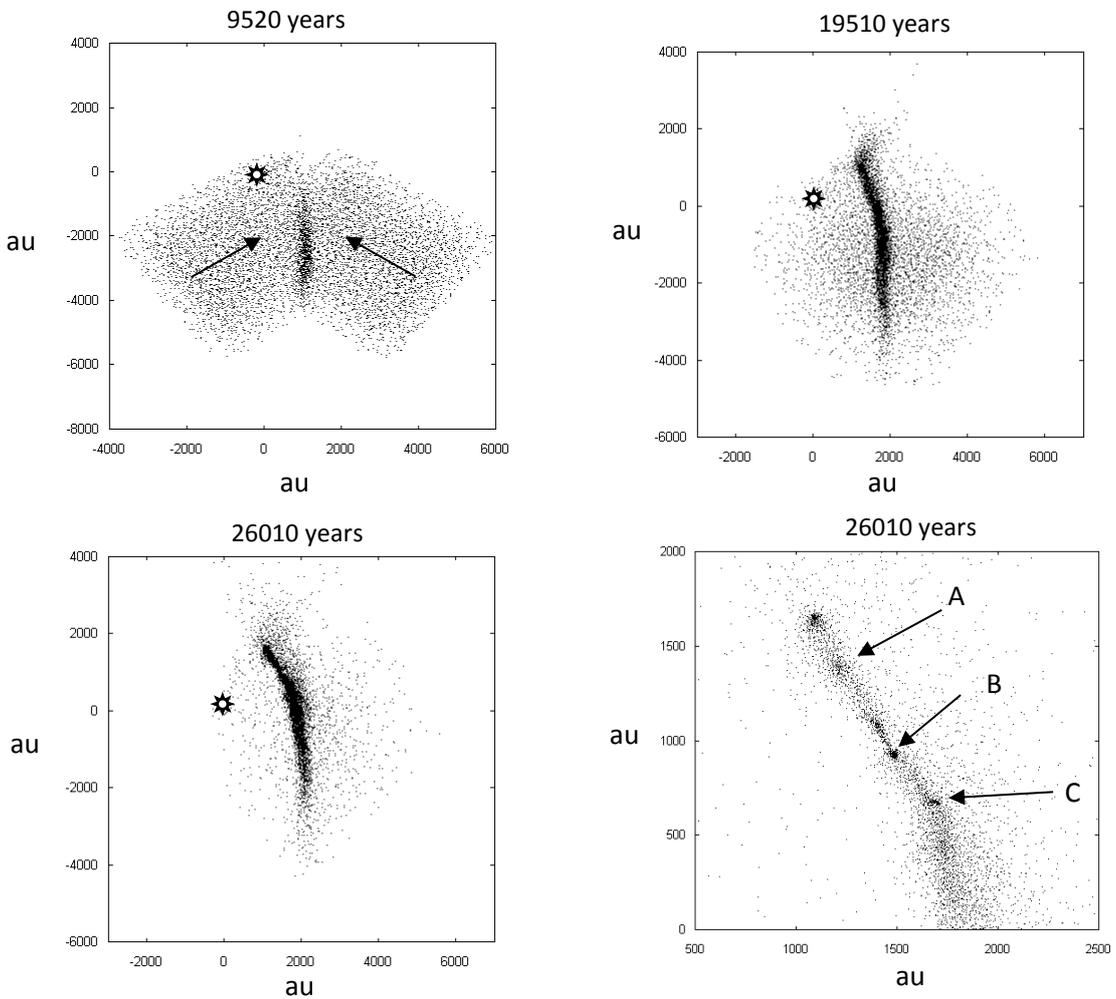

Figure 4 Simulation of colliding gas streams at times 9,520 years, 19,510 years and 26,010 years. The final frame shows a higher resolution view at 26,010 years (Oxley and Woolfson, 2004).

## 4 Angular momentum in the Solar System and models of planet formation

The first nebula-based theory for solar-system formation was given by Laplace (1796). A collapsing nebula evolved into a central core, which formed the Sun, and a disk within which planets formed. While this was the standard model for more than fifty years, by the middle of the 19th Century it was abandoned because it failed to explain the distribution of angular momentum in the Solar System. The Sun, with 99.87% of the mass of the system had only 0.5% of its angular momentum in its spin, the remainder being in the planetary orbits. No way of partitioning mass and angular momentum in that way could be envisaged at that time. The modern NT invokes a combination of mechanical forces (Lynden-Bell and Pringle, 1974) and some form of magnetic linkage between a central collapsing core and a surrounding disk (e.g. Armitage and Clarke, 1996) to transfer angular momentum from inner to outer material.

For the CT, which is a dualistic theory, the slow rotation of the Sun comes about by the process described in Section 2 and the orbital angular momentum of the planets comes from the motion of the protostar relative to the Sun. Indeed, because of their extreme nature, the initial planetary orbits contain too much angular momentum and we must address the problem of how planetary orbits evolve to their present states.

Another interesting difference between the two theories is the process by which planets are formed. The NT process is termed 'bottom-up', involving four stages in which objects of increasing size are produced. First dust in the disk settles into the mean plane, then gravitational instability in the so-formed dust carpet leads to the production of solid bodies, called *planetesimals*, of dimensions from hundreds of metres to tens of kilometres. The planetesimals aggregate to form either terrestrial planets or the cores of major planets and, finally, disk gas is accumulated by major planets to form extensive atmospheres. The timescale for this multi-stage process is of order $10^6$ years in the terrestrial region of the Solar System but increases rapidly with increasing distance from the Sun.

By contrast, the capture-theory process is called 'top-down' in which major planets are produced directly from an above-Jeans-critical-mass of dusty gas – the condensations seen in Figures 2, 3 and 4. To explain smaller bodies it is necessary for one or more of the initial planets to break up, a proposal for which is given in Section 9. As will be seen from Figures 2 and 4 the timescale for planet formation by the CT is of order $10^4$ years.

## 5 Evolution of orbits

Although difficult to see in Figure 2, but just discernible in Figure 4, some protostar material forms a disk around the star, usually of mass 25-50 $M_J$ but sometimes outside that range. The areal density and extent of the disk is variable but it usually extends out to several hundred au from the star. Sometimes the density fall-off outwards is in a quasi-exponential form but it can also be of a doughnut form with maximum density a few hundred au from the star, as shown in a CT simulation in Figure 5 in which the disk particles have been enhanced for clarity. The distance of the peak areal density is about 350 au from the star.

These disks form a resisting medium within which the planet moves. Before describing how orbits are modified we should say something about the geometry of a disk. The word 'disk' conjures up an image of a circular object of uniform thickness, like a coin, but a disk in quasi-equilibrium round a star would not be stable in that configuration. The areal density falls off to zero at the edge of the disk but in cross-section it has to flare out in a direction perpendicular to the disk mean-plane. The general appearance of a stable disk, modelled by a distribution of points, is shown in Figure 6.

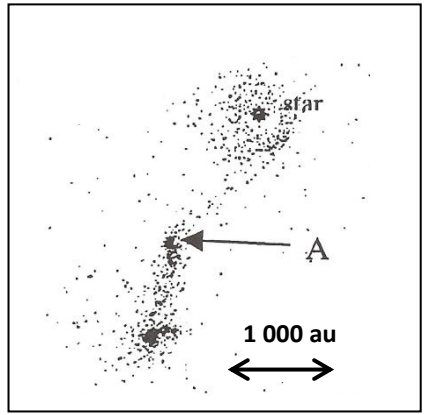

Figure 5   A capture-theory simulation showing a strong doughnut-like captured medium (Woolfson, 2003)

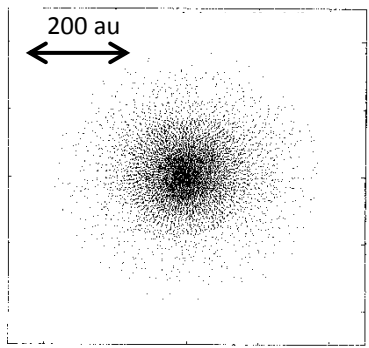

(a)

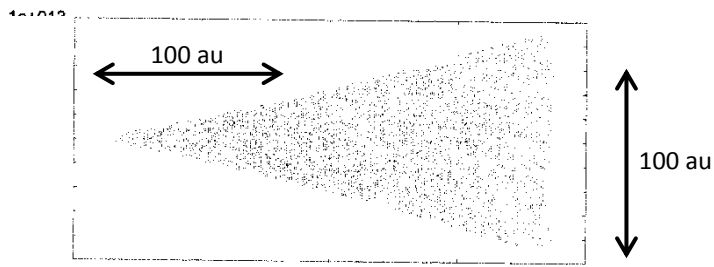

(b)

Figure 6   (a) Resisting-medium particles seen in plan-view.  (b)  The distribution perpendicular to the plane (Woolfson, 2003)

   The evolution of planetary orbits was described by Woolfson (2003) using a medium modelled as shown in Figure 6. The points were distributed so as to be most densely packed close to the star with density falling off with distance. The masses of the individual points were then fixed to give the required density distribution. The medium points did not act gravitationally with each other and were set in motion in Keplerian orbits around the central star so, in the absence of a planet, the medium was stable. The planet was then inserted into the medium with an initial orbit corresponding to what might be expected from a CT origin. The simulation was then run with gravitational interactions involving the star, planet and medium points, with the previously-given exception that medium points do not gravitationally affect each other.  From observations of disks around young stars, detected by the presence of an infrared bump added to the normal stellar emission, it is deduced that disk lifetimes are normally a few, typically three, million years but up to ten million years. For that reason, as the simulation is run, the mass of each of the medium points varies as

$$m(t) = m(0)\exp(-\gamma t) \qquad (1)$$

where $m(t)$ is the mass of a point at time $t$, giving an exponential decline in the density of the medium with time. The areal density of the medium, corresponding to a doughnut form in this case, is given by

$$\rho(r) = C\exp\{-\alpha^2(r-r_p)^2\} \qquad (2)$$

where $C$ is adjusted to give the total mass of the medium and this areal mass is spread in the direction perpendicular to the disk uniformly between $W$ and $-W$ where

$$W = \frac{\pi c}{2\omega} \qquad (3)$$

in which $c$ is the sound speed in the medium and $\omega$ is the local Keplerian angular velocity. Figure 7 shows the result of one calculation with this model. The parameters are:

Mass of planet = $4M_J$ ; initial semi-major axis = 1,500 au ; initial eccentricity = 0.9; mean molecular mass for medium = $2 \times 10^{-27}$ kg ; temperature of medium = 20 K; total medium mass = 50 $M_J$ ; star mass $M_\odot$; with $r$, the distance from the star in au, $\alpha = 0.007587$ au$^{-1}$; $r_p$ = 200 au; with time in years $\gamma = 10^{-6}$ year$^{-1}$; medium represented by 77,408 particles.

After 1.5 million years the orbit rounded-off and after eight million years the orbit decayed to a semi-major axis of about 2.4 au. Running with different parameters shows some obvious characteristics. For example, a more massive medium, or one that is longer-lasting, makes round-off and decay faster and increases the total orbital decay. An interesting feature of the orbital evolution is that the perihelion, $q$, steadily increases until round-off after which the circular orbit continues to decay.

Another result is that for a given medium the decay *increases* with the mass of the planet. In Table 1 the final semi-major axes, $a$, and eccentricities, $e$, are given for protoplanets of different masses with initial orbital parameters ($a$, $e$) = (1,500 au, 0.90). The medium was of Gaussian form as given by (2)

For some simulations, with dense, long-lasting media, the final semi-major axis is so small (making the calculation difficult to perform with tiny timesteps) that the conclusion is that the planet would plunge into the star. There are exoplanets with small semi-major axes – down to 0.015 au – and there exists a mechanism that can maintain that order of distance even when the medium is still present. It is the same mechanism that causes the Moon to retreat slowly from the Earth and depends on the spin period of the central body being less than the orbital period of the secondary body. The tide raised on the central body by the orbiting body is dragged forward by the central-body spin. The near tidal bulge on the central body then exerts a gravitational pull on the orbiter in the direction of its motion thus increasing its orbital angular momentum, which moves it outward. In the case we are considering, if the energy gain by the tidal interaction equals the energy loss due to the resistance of the medium then the orbit stabilizes.

Table 1  The variation of the final *a* and *e* with protoplanet mass for a given medium

| Mass of planet | Semi-major axis (au) | Eccentricity |
|---|---|---|
| $M_J$ | 2.980 | 0.0067 |
| $2M_J$ | 1.549 | 0.0065 |
| $3M_J$ | 1.056 | 0.0064 |
| $4M_J$ | 0.787 | 0.0064 |
| $5M_J$ | 0.633 | 0.0062 |

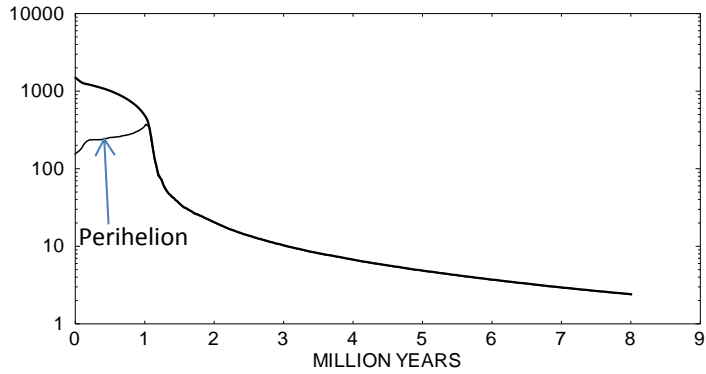

(a)

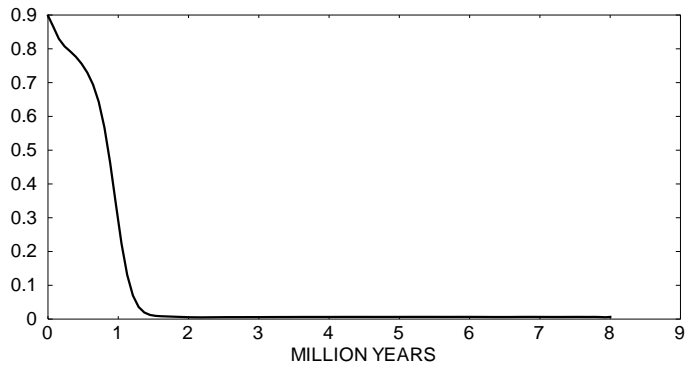

(b)

Figure 7.  The variation with time of (a) semi-major axis and perihelion and (b) eccentricity.

Sometimes the medium can be of low mass and extremely diffuse and/or the star can be very active with a strong stellar wind that quickly disperses the medium. In such cases the final semi-major axis of the orbit may be very large. There are some exoplanets, which can be directly imaged, that are at large distances from their stars, the largest known being that of the planet of mass 11 $M_J$ at distance 650 au from the star HD 106906 (Bailey et al., 2014). The current speculation concerning a possible solar-system Planet 9 with estimated (*a*, *e*) = (700 au, 0.6) could correspond to a protoplanet produced in an extremely extended heliocentric orbit that only partially decayed and rounded-off before the resisting medium was lost.

For a planet in an eccentric orbit moving in a medium the elements of which are in circular Keplerian orbits, the outcome is almost certainly a near-circular orbit. Although orbital evolution is occurring throughout the planet's orbit we can determine the general pattern of evolution by just considering what happens at periastron (closest distance to the star) and apastron (furthest distance from the star). At periastron the planet moves faster than the medium and so is slowed down. This keeps the periastron constant but reduces the apastron, thus reducing the eccentricity. At apastron the planet moves slower than the medium and so is speeded up. This keeps the apastron constant but increases the periastron (see Figure 7a), thus reducing the eccentricity. At both extremes the eccentricity decreases and if, as most commonly occurs, the periastron effect is stronger then the orbit will also decay. However, there are some exoplanet orbits of high eccentricity, up to 0.97, and we must consider how these could occur.

Young stars can be very active, in particular with strong stellar winds. The effect of these on the medium is to exert an outward force on it that neutralizes, to a greater or lesser extent, the gravitational influence of the star. The planet is virtually unaffected by the stellar wind but the medium is orbiting more slowly as though the mass of the star was reduced. At periastron the planet is still moving faster than the medium and so the apastron and eccentricity are reduced. With the medium slowed down it can now happen that the planet is moving *faster* than the medium at apastron and is slowed down and the effect now is that the periastron is reduced and the eccentricity *increased*. If the density at periastron is much greater than at apastron then the effect there will dominate and the orbit will still round-off, albeit at a slower rate. However, consider that the medium had the doughnut structure shown in Figure 5. The orbit would still initially decay and round-off but once the apastron reached the region of peak density the effect there, and going inwards from the peak, would dominate and the eccentricity thereafter increase.

Simulations were carried out for a planet of mass $5M_J$ with initial orbit $(a, e) = (1,500$ au, $0.9)$. The medium, of mass $50\ M_J$, had the form shown in (2). The simulations were run with different diminutions of the effective stellar mass for the medium. For effective stellar mass greater than 0.5 of the true mass the orbits are circular but for lesser values the final eccentricity steadily increases as the effective mass decreases. Figure 8 shows the variation of semi-major axis and eccentricity with time for three runs with different effective masses..

As another example of orbital evolution we consider exoplanets which have been discovered associated with binary stars, either orbiting one or both of them. A particular example is the binary system, γ-Cephei with stars of mass $1.40\ M_\odot$ and $0.40\ M_\odot$ with orbital parameters $(a, e) = (20.2$ au, $0.41)$, where an exoplanet orbits the more massive star with orbital parameters $(a, e) = (2.04$ au, $0.115)$ (Observatoire de Paris, 2004). For the CT the distance of the filament from the star is normally several hundred au, so a close-binary pair would act like a single gravitational centre of slightly varying strength and direction and the process of planet formation would resemble that for a single star. Although the captured protostar material forming a disk around both stars would be stirred up by the binary motions this would not affect the general form of the orbital decay of the protoplanet. As it approached the stars one of at least three things could happen to it. It could be left in an orbit around both stars, acquire enough energy from the pair of stars to be expelled from the binary system or be captured by one of the stars.

Thus far we have only dealt with single planets in a resisting medium but for the Solar System and some exoplanet systems there are several planets present. When orbits round-off and decay, protoplanets are influenced not only by the star and the medium but also by each other. The ratios of the orbital periods of pairs of the major planets of the Solar System are very close to the ratio of small integers, e.g.

$$\frac{\text{Orbital period of Saturn}}{\text{Orbital period of Jupiter}} = \frac{29.46\ \text{years}}{11.86\ \text{years}} = 2.48 \approx \frac{5}{2}$$

$$\text{and} \quad \frac{\text{Orbital period of Neptune}}{\text{Orbital period of Uranus}} = \frac{164.8 \text{ years}}{84.02 \text{ years}} = 1.96 \approx \frac{2}{1}$$

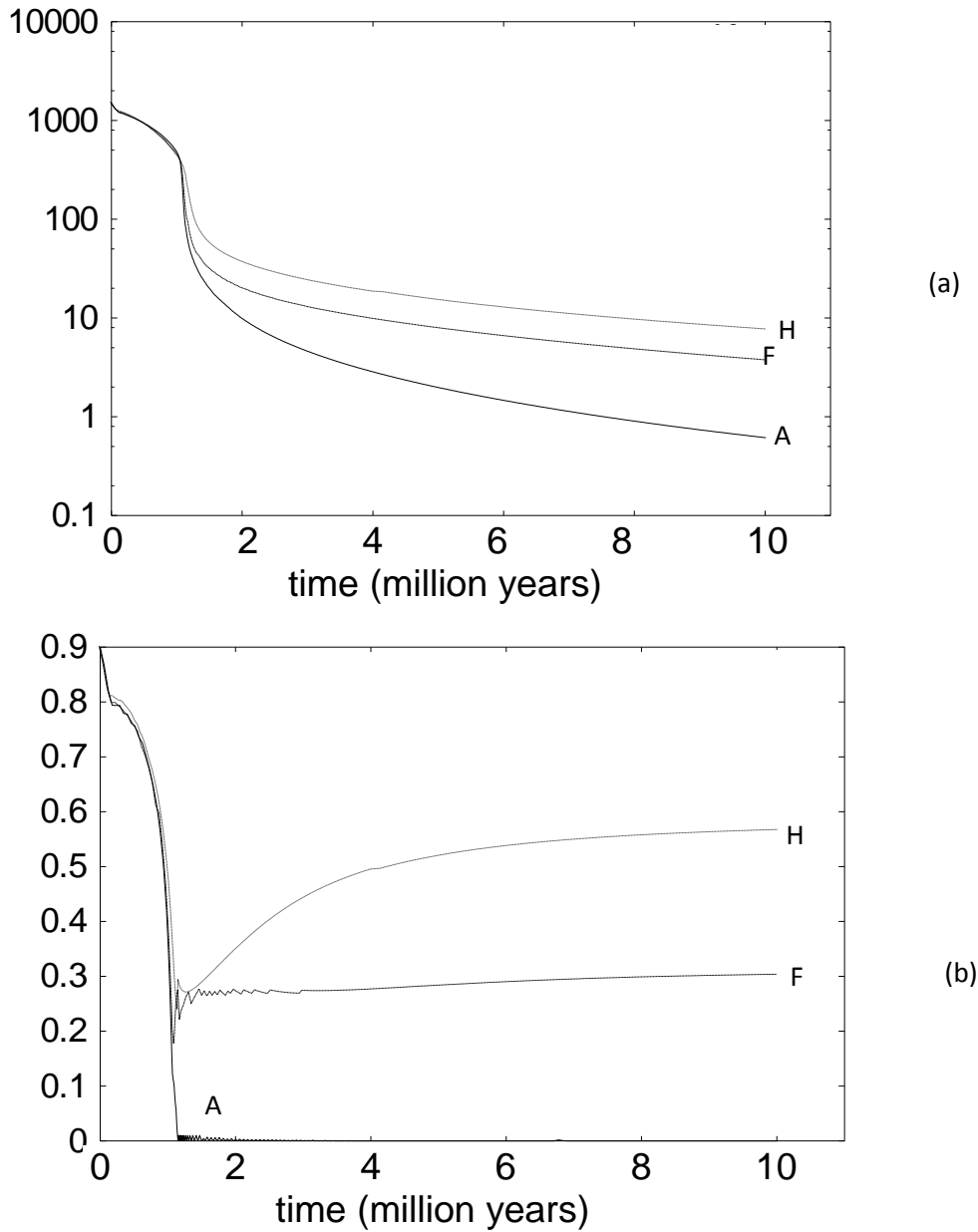

Figure 8   Three simulations of orbital evolution, two of which give eccentric orbits. (Woolfson, 2003)

Melita & Woolfson (1996) described a mechanism that explained how these orbital commensurabilities were established. For one computational model two bodies were placed in orbit, the inner one with the mass of Jupiter and the outer with the mass of Saturn. The initial eccentricities and inclinations were the same for both orbits – 0.1 and 0.06 radians (3.4°) respectively. The initial ratio of the orbital periods was set at 2.5, close to the present ratio for Saturn-Jupiter. Three runs of the computation were made with different initial relative positions in their respective orbits. Both the semi-major axes and the inclinations fall monotonically but the eccentricity falls at first but then rises again for both bodies. However, as seen in Figure 9, the ratio of the two periods departs from 2.5 and settles down close to 2.0, actually oscillating about 2.02, although both orbits are still decaying.

As with many numerical studies of complex systems, the computational results are easier to interpret than to predict in advance. When the computation is begun the eccentricities fall quite quickly, which reduce the speeds of the planets relative to the resisting medium. This reduces the rate of energy dissipation and hence the rate of change of orbital parameters. We have previously indicated that the rate of orbital decay increases with increasing mass of the planet but this is true if we consider only the so-called Type II migration mechanism (D'Angelo and Lubow, 2010)  This applies to planets of about Jupiter mass or more, where the planet clears a path in the resisting medium. Saturn's orbit decay is by the more effective Type-I migration mechanism (Lubow and Ida, 2011), where the body stays in contact with the medium, and so, although it is less massive than Jupiter, its orbit decays more rapidly than that of Jupiter. The differential decay changes the ratio of the periods of the orbits until it becomes close to 2.0. Now a new mechanism comes into play. We know that Kirkwood gaps in the asteroid belt correspond to periods commensurate with that of Jupiter. The reason for this is that Jupiter *removes* energy from *interior* bodies in commensurate orbits. Conversely, although there is no obvious demonstration of this in the Solar System, energy is *added* to *exterior* bodies in commensurate orbits. When commensurability is established the bodies come closest together repeatedly at the same points of their orbits. This amplifies the perturbations by a resonance effect and increases the eccentricity which, in its turn, increases the rate of dissipation and hence the rate of decline in the semi-major axis. The increasing dissipation due to the increasing eccentricity of Saturn's orbit is balanced by a gain of energy due to the 2:1 resonance. Thus the rate of change of Saturn's semi-major axis is less affected while that of Jupiter is increased because the effect of Saturn as an exterior body is to add to the loss of energy due to dissipation. The resonance state is now lost and once again Saturn decays more rapidly until resonance is re-established. The result of this is that the ratio of the periods, which had fallen from 5:2 to 2:1, now becomes locked at close to 2:1, with a slight variation, although both orbits are still decaying.

The numerical experiments were repeated for the Jupiter-Saturn system with different initial eccentricities – the nine combinations with each eccentricity having the possible values 0.1, 0.2 and 0.3 – but always with an initial orbital period ratio 2.5.   The results are shown in Figure 10. Five of the combinations end up with a ratio close to 2.0 and four stay at the original ratio 2.5.

Other trials gave a Neptune-Uranus system with a ratio close to 2.0 starting with the present observed ratio of 1.96 (Figure 11). A factor that has not been taken into account in the calculations is the evaporation of the resisting medium. If it effectively disappears before commensurability is established then some pairs of orbital periods may become stranded away from the commensurabilities towards which they were evolving. Thus if the orbit of Uranus was decaying more slowly than that of Saturn then the ratio of the periods would gradually increase. The present ratio of the periods is 2.85, which might indicate that it was going towards 3.0 but was terminated by the removal of the medium.

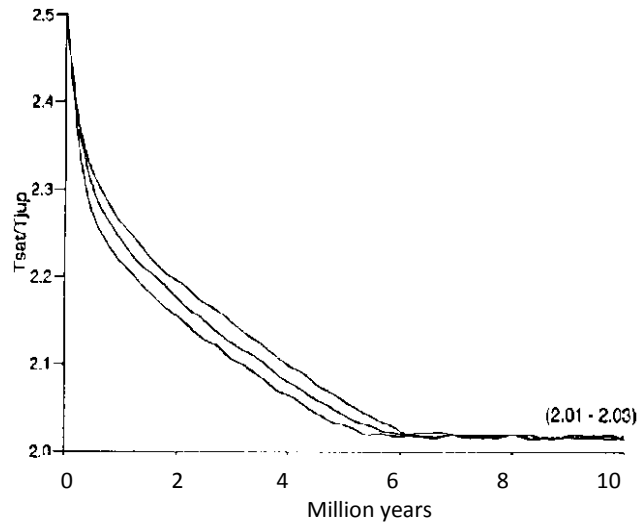

Figure 9 Ratio of periods for Saturn:Jupiter starting near the 5:2 resonance. Each curve represents a different initial relative position (Melita and Woolfson, 1996).

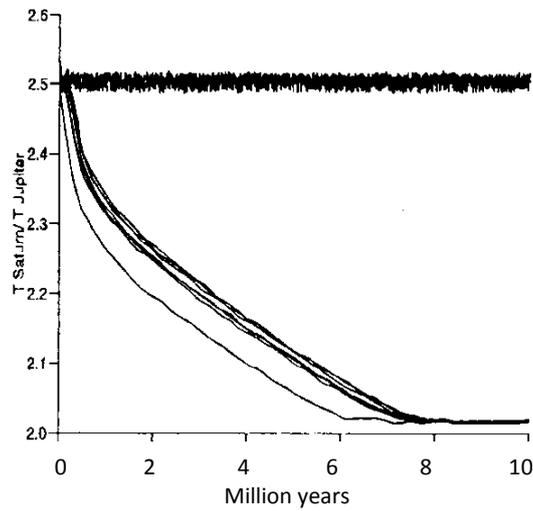

Figure 10 Evolution of the Jupiter-Saturn system for nine different pairs of initial eccentricities (Melita and Woolfson, 1996).

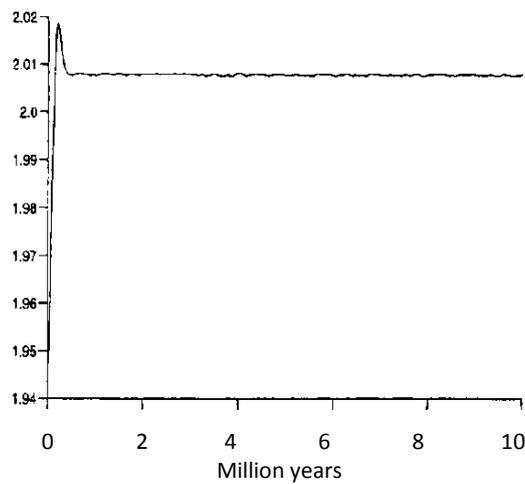

Figure 11 The Uranus–Neptune system starting near the present ratio of periods. The system reaches a resonant configuration with a ratio of periods slightly greater than 2 (Melita and Woolfson, 1996).

## 6 The formation of satellites

When Galileo first saw the large satellites of Jupiter through his telescope he interpreted it as a small-scale version of the Solar System and it reinforced his belief in the Copernican model. For him, and many that followed, it became axiomatic that the mechanism for producing satellites should be a small-scale version of that for producing planets.

The distribution of angular momentum is an important consideration in theoretical aspects of solar-system origin and development and here we compare the planetary system and satellite systems in terms of where angular momentum resides in them. What we do is to consider the following ratio with respect to a number of primary and secondary pairs of bodies. With 'intrinsic' meaning 'per unit mass' this is

$$S = \frac{\text{intrinsic orbital angular momentum of secondary}}{\text{intrinsic angular momentum for material at primary equator}} \quad (4)$$

This quantity is given for various pairs of bodies in Table 2, which clearly shows a distinct difference between the planetary system and satellite systems. The orbital angular momentum of satellite systems does not dominate to the extent that it does for the solar-system planetary system, a factor that led to the demise of the Laplace model.

Table 2 The ratio, *S*, of the intrinsic angular momentum of the secondary orbit to that of the spin of the central body at its equator

| Central body | Secondary body | Ratio *S* |
|---|---|---|
| Sun | Jupiter | 7 800 |
| Sun | Neptune | 18 700 |
| Jupiter | Io | 8 |
| Jupiter | Callisto | 17 |
| Saturn | Titan | 11 |
| Uranus | Oberon | 21 |

Figure 3 shows that as a protoplanet collapses a disk is left behind and it is found that the mass of the disk is comparable to that of the central collapsing core. Woolfson (2004a) described a process of satellite formation in the disk that exactly parallels the process of planet formation proposed for the NT. The steps are:

(i) Dust settles into the mean plane of the disk to form a dense dust carpet
(ii) The dust carpet is gravitationally unstable and fragments to produce solid bodies – satellitesimals.
(iii) The satellitesimals aggregate to form satellites

*Dust settling*

Although most dust particles in molecular clouds are of submicron size, which are easiest to detect because of the way they scatter and absorb light, there is a distribution of sizes

$$n(D) = KD^{-3.5} \qquad , \qquad (5)$$

where n(*D*) is the number density of particles with diameter *D* and *K* is a constant. Recent work suggests that particle diameters up to 5µm are present and Wood et al (2001) have detected dust particles up to 50µm in size in the dust disk surrounding a T-Tauti star. More recently, millimetre size grains have been detected in some disks. Here, for a circumplanetary disk, we accept an upper limit of 5µm and almost one-half of the mass is contained in particles between 2 µm and 5 µm. The larger particles move most rapidly towards the mean plane of the disk and as they do so they sweep up smaller particles.

The disk will have a flared structure, as shown in Figure 6b and an areal density decreasing outwards from the planet. Particles closer in will have less far to fall but do so in a denser medium that offers more resistance to motion. Settling times at different distances from a planet are shown in Figure 12, using the theory developed by Weidenshilling, Donn and Meakin (1989) for the following set of parameters

Mass of planet = $2.0 \times 10^{27}$ kg (approximately the mass of Jupiter)
Mass of disk = $2.0 \times 10^{27}$ kg
Areal density fall by factor *e* every $10^8$ km distance from planet
Mean molecular mass of gaseous material = $4 \times 10^{-27}$ kg
Temperature of disk material = 20 K
Density of dust grains = $3 \times 10^3$ kg m$^{-3}$
Ratio of principle specific heats of gas = 5/3

At distances beyond $4.0 \times 10^6$ km, for a reasonable disk lifetime of three million years, settlement would be complete. However, the total mass of the solid material in the carpet must be sufficient to provide the material for satellites. If we take the case of Jupiter the total mass of its large four satellites – the *Galilean satellites* – is about $4 \times 10^{23}$ kg. For the disk we have specified, the total mass beyond $4 \times 10^6$ km is over $8 \times 10^{26}$ kg and if just 0.5% of that is deposited dust then the total mass of the carpet would be $4 \times 10^{24}$ kg – an ample source of solid material to form the satellites. If larger dust particles were accepted then timescales could be much reduced.

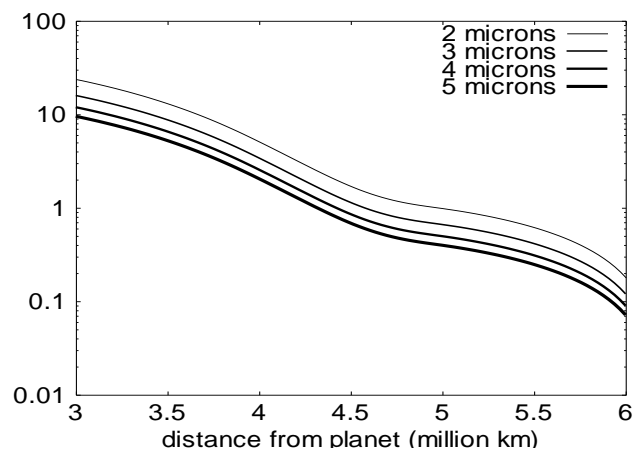

Figure 12 The settling time for dust at different distances from the planet and various particle diameters

*Formation of satellitesimals*

The theory for the gravitational instability of a dust carpet was given by Goldreich and Ward (1973). A necessary condition for a satellitesimal to form at any location is that it must be able to withstand disruption due to planetary-produced tidal effects. This is when

$$\rho_B \geq \frac{3M_P}{2\pi R^3} \qquad (6)$$

in which $\rho_B$ is the density of the spherical blob, $M_P$ the mass of the planet and $R$ the distance to the centre of the planet. Once the dust carpet reaches a thickness, $h$, such that its density reaches the critical level indicated by (6) then gravitational instability will set in and the carpet will begin to break up. If the areal density of the dust component of the disk is $\rho_{ad}$, then that thickness is given by

$$h = \frac{\rho_{ad}}{\rho_B} \; . \qquad (7)$$

According to Safronov (1972) the area of the disk in each of the condensations will be about 60 $h^2$ so the total volume of a condensation, which will form a satellitesimal, is about 60 $h^3$ with a mass, $m_B = 60\ h^3 \rho_B$. For the parameters that gave Figure 12 the masses of satellitesimals as a function of distance from the planet are shown in Figure 13.

Jupiter has 63 satellites, dominated by the Galilean satellites, Io, Europa, Ganymede and Callisto, with masses $8.93 \times 10^{22}$, $4.88 \times 10^{22}$, $1.497 \times 10^{23}$ and $1.068 \times 10^{23}$ kg respectively. At a distance of $4.0 \times 10^6$ km, where settling times are within the expected disk lifetime, a satellitesimal mass is about $2 \times 10^{22}$ kg, between about one-half to one-seventh the mass of the Gallilean satellites. The next most massive satellite of Jupiter is Amalthea, closer in than the Galileans and with a mass of $2 \times 10^{18}$ kilograms, within the range of masses of satellitesimals in the figure.

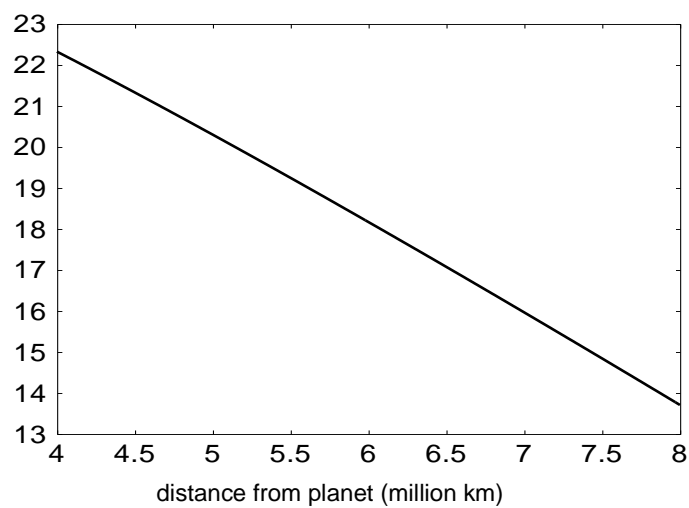

Figure 13   Satellitesimal masses at various distances from the planet

*From satellitesimals to satellites*

The expression found by Safronov (1972) for the time to form a planet, adapted to satellite formation, is

$$\tau_S = \frac{4 r_L \rho_m P}{3 \pi \rho_{ad}(1+\beta)} \tag{8}$$

where $r_L$ is the radius of the satellite, $P$ is the period of a circular satellite orbit in the region of formation, $\beta$ is a constant somewhere in the range 4 to 10, $\rho_m$ is the density of the body being formed and $\rho_{ad}$ is the mean areal density of satellitesimals. For a satellite of mass $10^{23}$ kg and density $2.5 \times 10^3$ kg m$^{-3}$, out to a distance $4.0 \times 10^6$ km all formation times are less than 100,000 years.

We found that for the model disk being considered it was necessary to go out to a distance of at least $4.0 \times 10^6$ km to have a dust carpet formed in a sufficiently short time but the orbital radius of Callisto, the outermost Galilean satellite, is $1.88 \times 10^6$ km. The answer is that protosatellite orbits are constantly decaying by the Type-I migration mechanism, as they grow and will still continue to decay for the duration of the planetary disk. Figure 14 shows the decay of a partially formed satellite, with constant mass $10^{22}$ kilograms, moving in the disk that gave Figure 12, but decaying so that the density falls everywhere by a factor of *e* every million years. Starting with a radius of $7.5 \times 10^6$ km the orbit decays to the orbit of Callisto, the outermost Galilean satellite, in about $3 \times 10^5$ years and to the orbit of Io, the innermost one, in about $8 \times 10^5$ years. These short decay times are due to the relatively high areal density of the protoplanetary disk compared to that of the circumstellar disks described in Section 5. The orbit is close to circular for the whole of the decay period.

The commensurabilities found for the periods of pairs of satellite orbits of Jupiter and Saturn are due to coupling between satellites as described by Melita and Woolfson (1996) for planetary-orbit commensurabilities.

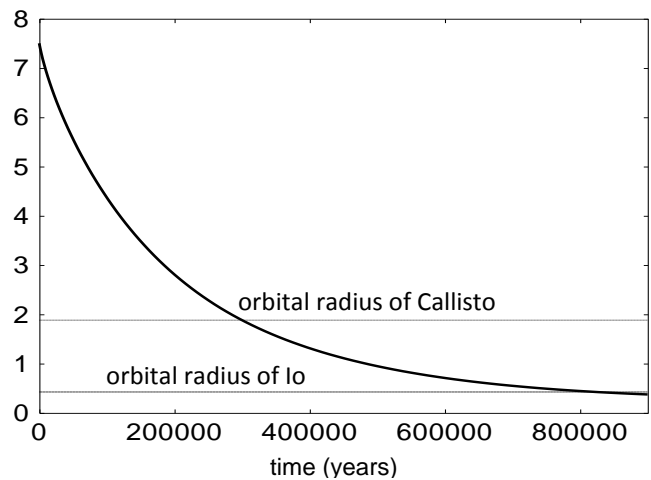

Figure 14  The decay of the orbit of a satellitesimal of mass $10^{22}$ kg.(After Woolfson, 2004a)

**7  The inclinations of exoplanet orbits**

In 2009 it was announced that the transiting exoplanet WASP-17b was in a retrograde orbit around its star (Anderson et al., 2010). Subsequently many more such orbits have been found from Kepler-mission observations using the Rossiter-McLaughlin effect (Rossiter 1924; McLaughlin, 1924), which involves observing the change in the peak wavelength of the emitted light as the transit progresses. Another, more recently-discovered, method depends on the measurement of blips on the transit intensity curve due to the presence of star spots (Nutzman, Fabrycky and Fortney, 2011). The histogram in Figure 15 is derived from spin-orbit misalignment (SOM, inclination) observations given by Heller (2013). It shows that retrograde orbits (inclination > 90$^o$) are not uncommon, although prograde orbits, especially with small inclinations, are predominant. In the latter category we can include the Solar System with average SOM 7$^o$.

For the CT there is a random relationship between the spin axis of the star and the plane of the star-protostar orbit, which defines the plane of the planetary orbits. This consideration alone would suggest that all inclinations should be possible and analysis shows that, assuming random inclinations, the probability of having an SOM $i$ is proportional to sin($i$), something that Figure 15 shows is not true.

The protostar material in the CT mechanism not only produces protoplanets and a resisting medium within which the protoplanets move, but also adds a small amount of material to the star. The angular momentum of this material pulls the spin axis of the star towards the normal to the exoplanet orbits and so reduces the SOM. Other material joining the star will be from the circumstellar disk, the inner part of which will drift inwards due to the mechanism described by Lynden-Bell and Pringle (1974). To give an idea of how much protostar material would need to be absorbed by a star to significantly affect the direction of its spin axis we note that one-third of a Jupiter mass in orbit at the solar equator has as much angular momentum as the Sun in its spin.

The final SOM depends on the initial value, due to the relationship of the star-protostar orbit to the stellar spin axis, and the mass of absorbed protostar material by the star. Since the latter could be up to a few Jupiter masses the likelihood of a small SOM could be high. The CT mechanism leads to a complete range of possible spin-orbit misalignments but with a strong bias towards small values (Woolfson, 2013a).

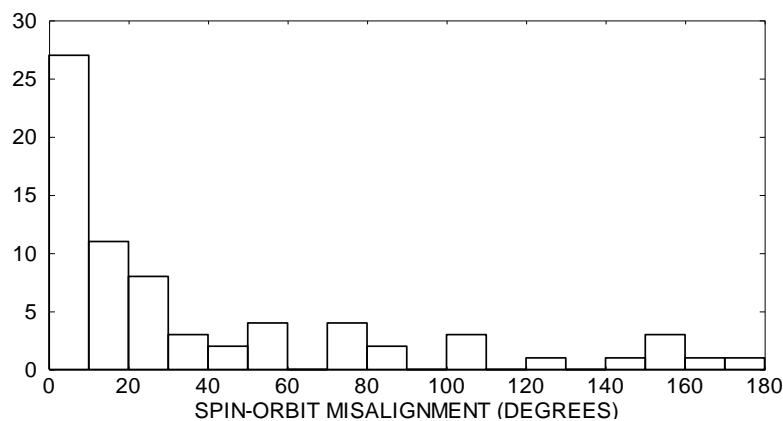

Figure 15  Spin-orbit misalignments (inclinations) for exoplanets (Woolfson, 2013b)

Again, it is interesting to contrast the CT and NT in their explanations for Figure 15. For the NT, the spins of the central star and disk would, in the absence of outside influence, be parallel and so be expected to give zero SOM. However, there are various factors that could change the initial SOM. Perturbation by an external body, such as another star, can produce the Kozai effect (Kozai, 1962) whereby there is an interplay between the eccentricity and inclination of an orbit and SOMs of up to 40° can be explained in this way (Fabrycky and Tremaine, 2007). Another effect of an external star is the possibility that the circumstellar disk, in which planets form, could either be changed in orientation or be warped by stellar perturbation. Because of different rates of orbital decay planets may interact closely and have the inclinations of their orbits greatly affected, even to the extent of complete reversal of motion into a retrograde orbit. Finally, there is a suggestion by Rogers, Lin and Lau (2012) that the effect of internal gravity waves can cause external layers of the star to spin round a different axis from the main bulk of the star. On this interpretation large SOM angles do not really occur. The spin of the great bulk of the star is perpendicular to the orbital plane of the planets. This effect has been simulated in two dimensions but it remains to be seen if three-dimensional simulations confirm this result.

The NT starts with a zero, or near-zero, SOM for each star and by various effects moves some of them away from zero. The CT begins with a wide range of SOMs and through absorption of protostar material by the star pushes them towards zero to different extents.

## 8 The proportion of stars with planets

Estimates of the proportion of stars with planets, based on observations, vary with time and tend to increase. A recent estimate is 0.34 (Borucki et al., 2011), although there are some estimates even higher.

It is clear from the description given at the end of Section 2 that interactions between protostars and condensed stars should occur frequently in the dense embedded state of a star-forming cloud and Woolfson (2016) developed a model to quantify the proportion of stars with planets. The model was investigated for SNDs of $n = 5,000$ pc$^{-3}$ to $25,000$ pc$^{-3}$ in steps of $5,000$ pc$^{-3}$ and initial protostar radii of $R_P = 1,000, 1,500$ and $2,000$ au with free-fall times of $t_{ff} = 10,200, 18,800$ and $28,900$ years, respectively, for a protostar of mass $0.3\ M_\odot$. As the protostar moves among the stars it collapses and its motion is followed for a maximum period $0.8 t_{ff}$, at which stage it has just over one half of its original radius.

A random configuration of $N_S$ stars is set up in a cubical cell, surrounded by 26 'ghost cells', each containing a similar configuration. A star moving out of the central cell re-enters on the opposite face, so keeping the density constant. The size of the cell, $a$, is set to give the required SND and stars in the ghost cells are only included out to a distance $ma$ from the centre of the central cell, so the spherical cluster contains, on average, $N_T$ stars where

$$N_T = 4\pi m^3 N_S / 3. \qquad (9)$$

We consider the motion of the protostar in a dense spherical fragment of the cloud as affected by stars plus the residual gas in the fragment. The gas is taken as uniformly distributed in a fragment and has two effects – firstly, gravitational and, secondly, by exerting a drag on bodies moving within it, so reducing their speeds.

Experience with CT calculations show that if the closest approach of the protostar orbit to a solar-mass star, $r_c$, was between 0.5 and 1.5 of the protostar radius then a CT event was almost certain to occur. Since the CT is based on tidal effects, for a star of mass $M_*$ this gives the condition

$$0.5\left(\frac{M_\odot}{M_*}\right)^{1/3} R_P \leq r_c \leq 1.5\left(\frac{M_\odot}{M_*}\right)^{1/3} R_P \qquad (10)$$

where $R_P$ is the current radius of the collapsing protostar

The masses of the stars are chosen by random selection from a distribution with mass index ~2.3. (Kroupa 2001), i.e.

$$f(M) \propto M^{-2.3}, \qquad (11)$$

with masses in the range 0.5 – 3.0 $M_\odot$, all greater than the mass of the protostar. This gives the average mass of the stars as 1.00 $M_\odot$.

For a collection of gravitationally-interacting stars in a bound region in equilibrium, and with no other forces acting, the virial theorem would be valid. This gives the kinetic energy, $K$, of the motions of the stars related to their total potential energy, $\Omega$, by

$$K = -0.5\Omega \qquad (12)$$

However, due to the action of gas-dynamical-friction, the stars in a dense embedded cluster will have sub-virial energy (Indulekha, 2013) so that

$$K = -\beta\Omega \qquad (13)$$

with $\beta < 0.5$. Protszkov et al. (2009) have considered $\beta$ in the range 0.04 to 0.15. This range of $\beta$ corresponds to root-mean-square speeds between 28% and 55% of the virial value. Stars, and the protostar, were given speeds corresponding to the equipartition value in a randomly-chosen direction. Thus with $N_T$ stars in the cluster the $i^{th}$ star has a speed

$$V_i = \left(-\frac{2\beta\Omega}{N_T M_i}\right)^{1/2}. \qquad (14)$$

The speeds assigned to stars of the extreme masses, 0.5 $M_\odot$ and 3.0 $M_\odot$ as a function of the SND are, with $\beta = 0.04$,

For $n = $ 5,000 pc$^{-3}$ and $M_i = 0.5\,M_\odot$  $V_i = 483$ m s$^{-1}$
For $n = $ 5,000 pc$^{-3}$ and $M_i = 3.0\,M_\odot$  $V_i = 198$ m s$^{-1}$

For $n = 25,000$ pc$^{-3}$ and $M_i = 0.5$ $M_\odot$    $V_i = 632$ m s$^{-1}$
For $n = 25,000$ pc$^{-3}$ and $M_i = 3.0$ $M_\odot$    $V_i = 258$ m s$^{-1}$

For comparison, Gaidos (1995) gave stellar speeds in a DEC in the range 500 m s$^{-1}$ to 2,000 m s$^{-1}$, corresponding to a larger value of $\beta$ and allowing larger values of $n$.

The equations of motion are numerically solved for the $N_S$ stars plus protostar in the basic cell. The gravitational effects of the stars in the ghost cells are taken into account but they are not moved during the integration step. The gravitational acceleration due to the gas, of mass $M_T$, within the fragment on a star at vector position **r** was included and is given by

$$\mathbf{a}_r = \frac{GM_T}{(ma)^3}\mathbf{r} \tag{15}$$

The integration is carried out using the 4-step Runge-Kutta method. If the motion of the protostar gives a periastron distance from a star between 0.5 and 1.5 of its current radius (which changes with time) then planet formation is deemed to have occurred but if it moves closer than 0.5 times its radius then it is taken that the protostar is disrupted without planet formation. If the planet formation or protoplanet disruption stage is reached then the trial is terminated and the next one begun, otherwise the calculation is terminated at time $0.8t_{ff}$. At the end of each integration step any stars, including the protostar, that have left the basic cell are reintroduced into the cell as previously described and the ghost cells and the cluster fragment are redefined. For four sets of conditions, with $\beta = 0.04$, 1,000 Monte Carlo trials were run. The sets were:

    A   $N_S = 8$, $m = 1.5$, giving $N_T = 113$
    B   $N_S = 23$, $m = 1.0$, giving $N_T = 96$
    C   $N_S = 10$, $m = 1.0$, giving $N_T = 42$

All these sets had the ratio of mass of gas to stars equal to 1. Set D is as for C but with the mass ratio equal to 3. The percentages of capture events is given in Table 3

A number of restrictive conditions were imposed on the analysis, which are:

(a)   Capture-theory events involving the collision of high-density regions have not been included.
(b)   The SNDs have been considered up to $2.5 \times 10^4$ pc$^{-3}$. Maximum values up to $10^6$ pc$^{-3}$, have been suggested by some authors (McCaughrean and Stauffer, 1994)
(c)   The maximum radius of a protostar has been taken as 2,000 au; other authors have suggested larger values.
(d)   The value taken for $\beta$, 0.04, gives root-mean-square speeds less than those indicated by Gaidos (1995). Increasing $\beta$ gives a higher proportion of stars with planets.

Since the initial orbits of the protoplanets are very extended then initial planetary systems can be disrupted by stellar perturbation in a dense embedded cluster before the orbits have completely evolved. This possibility has been analysed by Woolfson (2004b). It is concluded that up to 20% of planetary systems can be lost in this way but, nevertheless, the CT model seems capable of explaining the estimated proportion of stars with planets as deduced from present observations

Table 3  Percentage of CT interactions giving planets with various $R_P$ and $n$

| $R_p$ (au) | $n$ | Set A Capture (%) | Set B Capture (%) | Set C Capture (%) | Average Capture (%) | Set D Capture (%) |
|---|---|---|---|---|---|---|
| 1,000 | 5,000 | 0.7 | 0.6 | 0.3 | 0.5 | 0.3 |
|  | 10,000 | 1.8 | 1.6 | 1.0 | 1.5 | 0.8 |
|  | 15,000 | 3.4 | 2.1 | 1.8 | 2.4 | 1.9 |
|  | 20,000 | 3.8 | 3.3 | 2.9 | 3.3 | 3.2 |
|  | 25,000 | 5.3 | 4.5 | 4.6 | 4.8 | 5.3 |
| 1,500 | 5,000 | 2.3 | 2.3 | 2.7 | 2.4 | 2.4 |
|  | 10,000 | 6.5 | 5.6 | 5.9 | 6.0 | 5.5 |
|  | 15,000 | 11.6 | 8.9 | 10.2 | 10.2 | 11.0 |
|  | 20,000 | 11.5 | 13.5 | 11.9 | 12.3 | 12.9 |
|  | 25,000 | 15.9 | 16.3 | 17.1 | 16.4 | 19.5 |
| 2,000 | 5,000 | 7.9 | 6.6 | 5.0 | 6.5 | 5.2 |
|  | 10,000 | 17.7 | 16.9 | 16.7 | 17.1 | 17.0 |
|  | 15,000 | 29.0 | 25.9 | 24.7 | 26.5 | 29,1 |
|  | 20,000 | 36.4 | 37.8 | 33.6 | 35.9 | 40.8 |
|  | 25,000 | 48.0 | 46.0 | 44.2 | 46.1 | 52.4 |

The above analysis does not give the proportion of stars with planets but rather the proportion of *protostars giving planets*. Although not strictly valid, the following example gives an idea of how to transform from the latter quantity to the former. If 100 protostars give 25 sets of planets and 10 are disrupted then we have added 25 planetary systems (to pre-existing stars) and 65 new stars to the cluster. The ratio of the number of planetary systems produced to the number of stars added is then 0.25/0.65 = 0.38.

## 9  The larger terrestrial planets

The CT gives a plausible model for the formation of major planets, but it does not explain the terrestrial planets. Protoplanets form because the total mass of a blob in the filament, mostly gas, is usually greater than the Jeans critical mass. The initial movement of protoplanets is towards aphelion and by the time they approach the star closely they are compact objects capable of resisting disruption; gaseous exoplanets, so-called 'hot Jupiters' are observed at distances down to 0.015 au from their stars. This raises the problem of how the terrestrial planets were formed.

The idea of invoking a planetary collision to explain solar-system terrestrial planets was first suggested by Dormand and Woolfson (1977) but the state of knowledge and computational tools available at that time did not enable a realistic model to be created. When the protoplanet orbits were evolving, the mass of the resisting circumsolar disk gave a non-central force on the planets, causing orbital precession. Differential rates of precession meant that slightly inclined pairs of orbits would intersect in space from time to time during the orbital evolutionary period. Dormand and Woolfson postulated an initial system of six planets, the present four major planets plus two others, and found that the probability that some pair would collide before they all rounded off was of order 0.1 – small but not negligible. The idea that planetary collisions take place has received support from a NASA Spitzer Space Telescope observation in August 2009 of evidence of a planetary collision in the vicinity of the young star HD172555 (age 12 My) within the last few thousand years. In the light of new knowledge and the availability of suitable computational tools the planetary-collision hypothesis has been revisited (Woolfson, 2013b).

The distribution of deuterium in the colliding planets plays a significant role in the process. The cosmic D/H ratio is about $2 \times 10^{-5}$ but is much higher in many solar-system bodies. In star-forming clouds the *overall* D/H ratio is similar to the cosmic ratio but the ratio within the clouds is extremely non-uniform. The D/H ratio in some molecular species in these clouds, and in low-mass protostars formed within them, is considerably higher (Roueff et al., 2000: Loinard et al., 2001; Loinard et al., 2002; Parise et al., 2002). The average over all the molecular species is estimated at D/H > 0.01. This concentration of deuterium is due to the phenomenon of *grain-surface chemistry*. A deuterium atom falling on the surface of an icy grain will exchange places with a hydrogen atom in a molecule because this lowers the energy of the molecule and increases its stability. Over a long period this process concentrates the deuterium in ice molecules to the levels observed. Although the most common hydrogen-containing molecules are water, ammonia and methane, in cold clouds more complex molecules are present in considerable quantities; the ratio of methanol to water, $CH_3OH/H_2O$, has been found to be in the range 0.1 to 0.5.

A protoplanet substantially collapses in about $10^4$ years (Figure 3) and thereafter slowly evolves to its final state. The early stage of collapse, which is almost free-fall, is very slow and during this period solid grains will migrate towards the centre. Eventually an iron core with a silicate mantle would form, surrounded by a shell of now-vaporized hydrogen-containing molecules with a high D/H ratio. Over time, this excess deuterium would migrate outwards to increase the D/H ratio in the gaseous envelope.

Woolfson (2013b) postulated that the early Solar System contained six major planets, the existing four plus Enyo of mass 1.9 $M_J$ and Bellona of mass 2.5 $M_J$, both masses being well within the range observed for exoplanets. The planets were modelled in four layers based on incomplete settling of material by density – an iron core with some silicate, a silicate mantle, with some iron, a deuterium-rich gaseous shell with some silicate and a hydrogen-helium atmosphere. The overall composition is given in Table 4

Point-mass models for the SPH simulation were formulated, as described by Woolfson (2007), in which the density of SPH points was highest at the centre, the region of greatest interest for this simulation, and decreased with increasing distance from the centre. Figure 16 shows the projected model for the two planets just before the collision occurred. The four shells – gas, ice, mantle and core are shown alternately as black and white with each planet represented by 4,921 SPH points. A Tillotson equation-of-state (Tillotson, 1962) was used for the inner three regions and a modified gas law for the atmosphere, which accommodated the high pressure regions. Starting with the temperature and density at the centre given by Table 4, the equations for gravitational and pressure equilibrium were integrated outwards with the criteria that the density was discontinuous at boundaries between regions, where the equation of state changed, but that temperature was continuous. The boundary of the planet was taken when the temperature fell to 100 K, which gave the radii in Table 4.

Table 4   The characteristics of the colliding planets

| **Planet** | **Bellona** | **Enyo** |
|---|---|---|
| Mass ($M_\oplus$) | 798.75 | 598.37 |
| Radius (km) | $9.152 \times 10^4$ | $8.647 \times 10^4$ |
| Central density (kg m$^{-3}$) | 176,500 | 146,500 |
| Central temperature (K) | 85,000 | 74,000 |
| Mass of iron ($M_\oplus$) | 3.00 | 2.50 |
| Mass of silicate ($M_\oplus$) | 12.00 | 10.00 |
| Mass of ice ($M_\oplus$) | 6.00 | 5.00 |

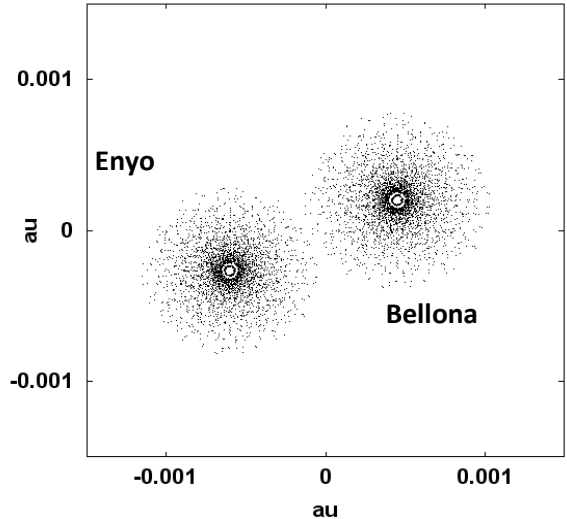

Figure.16  The four-layer structure of the colliding planets (Woolfson, 2013b)

The planets started apart from each other moving at such a speed that the relative speed at contact was 90 km s$^{-1}$. With respect to the centre-of-mass of the two-body system the planets were moving on parallel paths with an centre-to-centre offset of $7 \times 10^4$ km.

When the shock front from the collision reached the Enyo deuterium-rich region the temperature was at a level that gave a high rate of D-D nuclear reactions. Holden and Woolfson (1995) had examined in detail the nuclear reactions that would occur with a mixture of deuterium-rich ices and silicates. They found that that once the temperature reached $5 \times 10^8$ K the deuterium was exhausted and heat generation by other thermonuclear reactions was at a lower rate. In the present simulation, when an ice SPH particle reached a temperature of $3 \times 10^6$ K (when the rate of D-D nuclear reactions was high) the temperature was immediately raised to $4 \times 10^8$ K, lower than the Holden and Woolfson results indicated. This simplified the incorporation of the nuclear reactions without biasing the results towards exaggerating the effect of those reactions. The locations in which nuclear reactions took place quickly spread to other regions of D/H enhancement.

Figure 17 shows the progress of the simulation. By frame (c) nuclear reactions had occurred as seen by the rapid outward motion of material. This expansion became greater for successive frames but parts of the cores remained compact and steadily moved apart.

The approach speed of the planets when they were far apart, 42.9 km s$^{-1}$, suggests a collision in the terrestrial region. Since the colliding planets were the most massive in the initial planetary system, their orbits would have evolved most quickly by the Type-II migration mechanism and gone closest to the Sun (Table 1). It is proposed that the residual cores seen in Figure 17 formed the Earth and Venus, from the Bellona and Enyo residual cores, respectively. Their estimated masses from the simulation are 2.4 $M_\oplus$ and 1.5 $M_\oplus$, both too large, but the model indicates how the larger terrestrial planets could have originated.

Earth and Venus would have formed by this process while a resisting medium was still in place, although somewhat depleted, and, with a highly centralized medium, their orbits would have rounded off within the terrestrial region. `

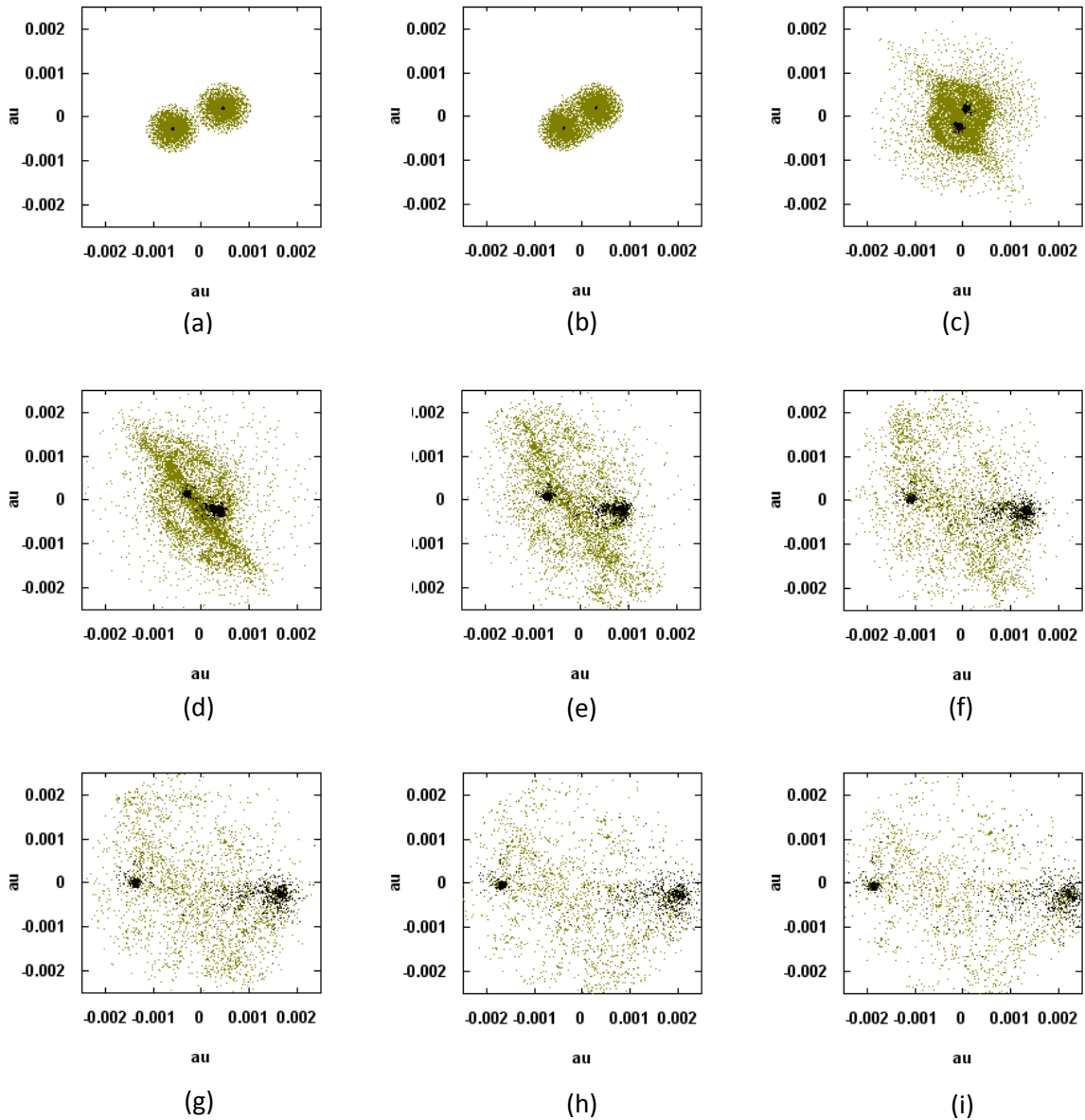

Figure 17  The progress of the collision. (a) $t = 0$, (b) $t = 590$ s, (c) $t = 1{,}326$ s, (d) $t = 2{,}505$ s (e) $t = 3{,}917$ s, (f) $t = 5{,}336$ s, (g) $t = 6{,}415$ s, (h) $t = 7{,}597$ s, (i) $t = 8{,}609$ s (Woolfson, 2013b) .

## 10  The Moon

Bellona and Enyo, like Jupiter and Saturn, would have had many satellites and, because their masses are both greater than that of Jupiter, some of their satellites could have been more massive than Ganymede. Following the collision, the possible outcomes for a particular satellite are limited to the following:

(I)     The satellite could be retained by one or other of the planet residual cores.
(II)    The satellite could end up in a heliocentric orbit.
(III)   The satellite could escape from the Solar System.
(IV)   The satellite could be completely disrupted by debris from the collision.

Trial calculations show that the most likely of these outcomes are (II) and (III). The least likely is (IV) – a massive amount of debris would be required to disrupt a substantial satellite although a small satellite might be disrupted in this way. We propose that the Moon was a satellite that was left in orbit around the Bellona residual core – now Earth. In both mass and density it is intermediate between the Galilean satellites Io and Europa, supporting the view that it was formed by the normal process described in Section 6.

In 1959 the Soviet Union Luna 3 spacecraft revealed the hemispherical asymmetry of the Moon. The nearside is dominated by large maria while the far side is predominantly heavily-cratered highlands. Altimeter measurements from a lunar satellite revealed that the far side had large basins, so the Moon had been bombarded uniformly by large projectiles, but the far-side basins had not filled with magma to give maria. The suggestion was made that this was due to a difference of crustal thickness on the two sides, which was confirmed when seismometers were left on the Moon by Apollo astronauts. While crustal thickness varied from place to place, the average thickness on the far side is about 12 km greater than that on the nearside.

An early satellite, formed as a fluid or plastic body in synchronous orbit around a planet, should have a *thicker* low-density crust on the nearside (Figure 18). The Moon formed around either Bellona or Enyo would have had a figure and internal distribution of material that would have ensured that when it orbited the residual core it would eventually have presented the same thicker-crust face to the Earth.

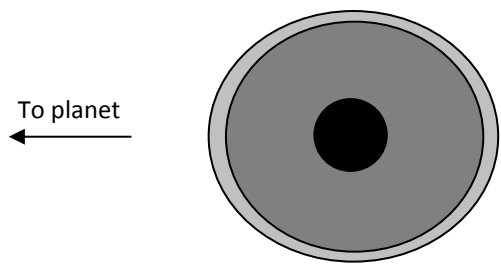

Figure 18  The initial structure of a satellite formed in synchronous orbit around a planet showing core (black), mantle (dark grey) and crust (light grey). The satellite distortion and thickness of the crust are exaggerated.

Due to near-surface convection, driven by solidified surface material sinking in the less dense liquid material below, the lunar crust would have solidified to some depth in the few million years between lunar formation and the planetary collision. The nearside of the Moon, facing the collision, was bombarded by debris travelling at about 100 km s$^{-1}$. Sharing the debris energy with lunar surface material, given that the escape speed from the Moon is 2.4 km s$^{-1}$, would have led to massive abrasion of the nearside and it has been estimated that up to 50 km thickness of surface material could have been removed in this way (Woolfson, 2013b). This would require that the arriving debris removed up to about eight times its own mass that, since the energy of the arriving debris has 1,600 times the intrinsic energy of escape from the Moon, is quite feasible.

The current most-supported model for the formation of the Earth-Moon system proposes that, shortly after solar-system formation, a Mars-mass body (Theia) struck the Earth obliquely and that, from the debris of this collision, the Moon accreted in orbit around the Earth. This mechanism has been realistically simulated by Benz, Slattery and Cameron (1986). An aspect of planetary science that receives a great deal of attention is that of attaching dates to the formation of the bodies it contains. This is usually done by radioactive dating but this method does not necessarily indicate when a body came into existence but rather when the material being examined became a closed system, i.e. solidified so that daughter products of radioactivity remained in the vicinity of the parent isotope. Meteorites are fragments of asteroids, which are usually assumed to be the original constituents of the Solar System and are small enough to have cooled quickly. The oldest meteorite age estimate, 4.567 billion years, is taken to be the age of the Solar System. Magma flows into mare basins occurred up to about three billion years ago and this material gives no age indication for lunar origin. However, highland rocks give ages up to 4.53 million years and the difference from 4,567 billion years is taken to indicate that the Theia-collision event occurred between 30 and 100 million years after solar-system formation. This deduction is based on solids forming on the Moon's surface shortly after its formation and remaining there until collected by Apollo astronauts.

Maria, plus the unfilled basins on the far side, cover about one-third of the Moon's surface Basins were produced by asteroid collisions and at their centres the magma depth can be 5-6 km (Head, 1976), indicating an original excavation depth of order 7-8 km. Making an extreme assumption that the average depth of maria excavation was only 1 km and that only 10% of the ejecta was retained and spread uniformly over the lunar surface, the average depth of cover in the highlands would be 30 m – probably a gross underestimate. The highland rocks from the Apollo 16 mission probably come from the nearby Mare Nectaris, the solidification ages of which would depend on their excavation depth. Any conclusions about the time of formation of the Moon based on dating highland rocks cannot possibly be valid. Another way of estimating the Moon's age depends on assumptions about the accretion of material on solar-system bodies after the Theia impact and is model dependent. It also gives extremely dubious age estimates for Mars and Venus (Jacobson et al., 2014).

## 11  Mars and Mercury

Mars and Mercury, with about four times and twice the mass of Ganymede respectively, are proposed as escaped satellites. The relatively high eccentricities of their orbits, 0.093 and 0.206 respectively, may indicate that the orbits of these small bodies evolved slowly and that their evolution was terminated before round-off by the complete loss of the circumsolar disk.

Like the Moon, Mars has hemispherical asymmetry with magma-covered northern plains and heavily-cratered southern highlands with one large deep depression, the Hellas Basin (Figure 19). A 2 km high scarp separates the two hemispheres. Heavy abrasion of one face of Mars could have removed much of the solid crustal material on the exposed side and volcanism would then have produced a magma plain. The Hellas Basin represents the effect of an exceptionally energetic projectile that penetrated the crust of the southern highlands.

Due to tidal coupling to the Earth, the Moon's spin axis is contained within its plane of asymmetry.. For Mars, which is not linked to another body, the plane of asymmetry makes an angle of 35° with the spin axis. Mars would have had a molten mantle over which the lithosphere could move – something akin to continental drift. A theorem by Lamy and Burns (1972) states that a spinning body with internal energy dissipation eventually settles down with its spin axis along the principal axis of maximum moment of inertia, a process known as *polar wander*. McConnell and Woolfson (1983) modelled the surface features of Mars either as positive features – raised regions in the highlands such as the Tharsis uplift, Argyre plain, Elysium plain and Olympus Mons – or as negative features such as the northern plains and the Hellas Basin. They calculated that the principal axis of maximum moment of inertia was 11.9° from the spin axis. The probability of being this close just by chance is about 0.02 and the discrepancy might be due to the crudeness in modelling surface features, or, perhaps, that polar wander was incomplete before the underlying mantle became too rigid to maintain it.

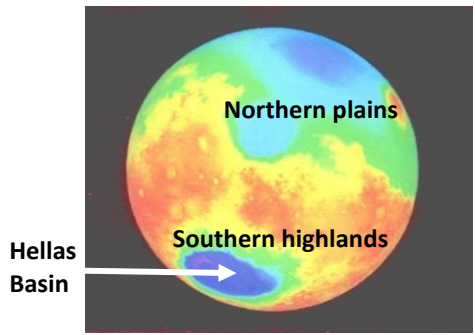

Fig.19  The topography of the Martian surface

The density of Mercury is higher than the uncompressed density of any other planet. Its iron core is similar in size to that of Mars and it has been suggested that it was once of similar size to Mars but had much of its mantle stripped away by a collision with another body. If Mercury had been a close satellite, or in the plane of the collision interface where the debris density was very high (Figure 17), then a large proportion of its mantle that faced the collision could have been stripped away. When it reorganized itself into a spherical body the motion of surface material, flowing round the surface and meeting in one small region, could have resulted in an overshoot with the following collapse giving the 'bullseye' feature, the Caloris Basin, and the diametrically opposite Chaotic Terrain, which regions are sub-solar at alternate perihelion passages (Woolfson, 2011a).

## 12  The Neptune-Pluto-Triton system

Pluto, a dwarf planet, has an orbit of eccentricity 0.249 and inclination $17^{o}$ that passes just within the orbit of Neptune, although the bodies never approach closely. Triton, the seventh largest satellite in the Solar System is in a retrograde orbit around Neptune, which rules it out as a regular satellite. Another peculiar satellite of Neptune is Nereid, in an extended direct orbit of eccentricity 0.75. Woolfson (1999) explained the relationship between these bodies as another outcome of the planetary collision.

The scenario giving this outcome is that Triton was a satellite of a colliding planet released into an extended heliocentric orbit taking it beyond Neptune. Neptune had a family of regular satellites the largest of which was Pluto with mass about two-thirds that of Triton. A computer simulation was made of a collision involving Triton and Pluto. The starting point is illustrated in Figure 20. Triton was in a direct heliocentric orbit with perihelion 2.6 au and aphelion 55.6 au.  Pluto was in a direct circular orbit, of radius 545,000 km, around Neptune. The before-and-after collision situations are shown in Figure 21. Triton, moving towards the Sun, strikes Pluto a glancing blow that ejects it from its orbit around Neptune into a heliocentric orbit with $(a, e)$ = (39.5 au, 0.253), very similar to its present orbit. It is set into retrograde spin and a large part of it is sheared off to form its comparatively large satellite Charon in a retrograde orbit – a process similar to the Benz et al. Moon-formation process. Fragments form its smaller satellites.

Triton loses energy in the collision and is captured by Neptune into a *retrograde* orbit with $(a, e)$ = (436,000 km, 0.88). Tidal effects give rapid round-off and decay of retrograde satellite orbits (McCord, 1966), so giving Triton's present orbit with $(a, e)$ = (355,000 au, 0.000).

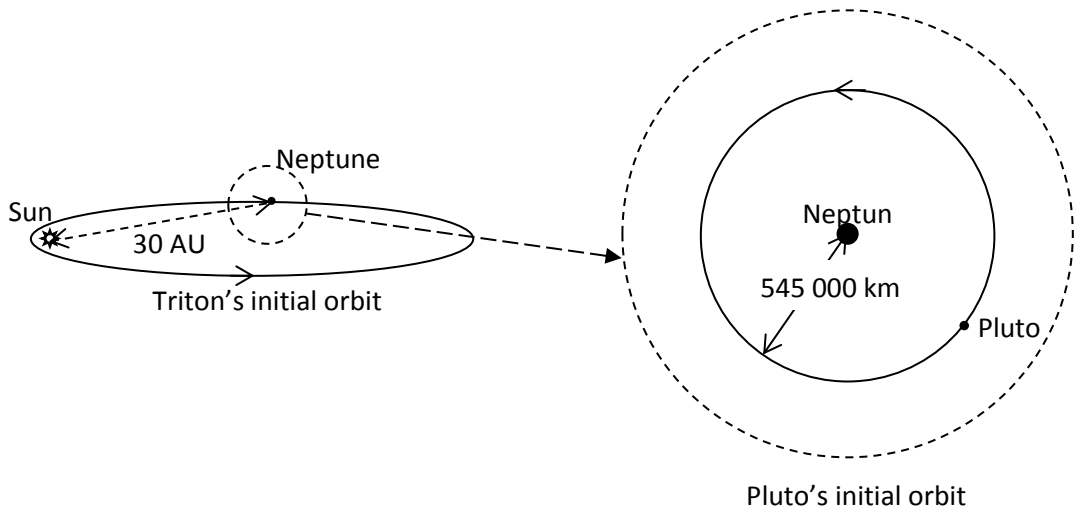

Figure 20  The initial orbits of Triton and Pluto before the collision.

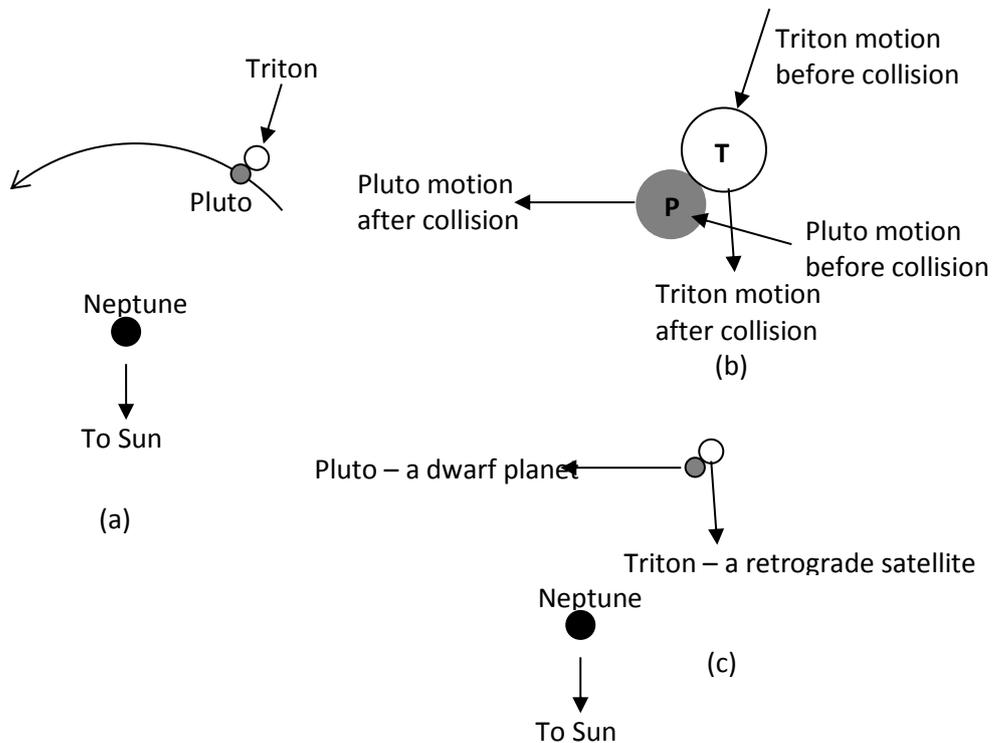

Figure 21    The Triton-Pluto collision.   (a) Triton, travelling towards the Sun strikes Pluto. (b) Motions before and after the collision.   (c) The final outcome.

The extreme nature of Nereid's orbit may be due to disturbance by Triton's incursion into the vicinity of Neptune. Alternatively, and perhaps more probably, Nereid could be a captured body – a large asteroid or small escaped satellite – that underwent a collision in the vicinity of Neptune, lost energy and was captured by the planet into its present highly-eccentric orbit.

## 13 Asteroids and comets

The residual cores of the colliding planets contained only a small part of the total inventory of iron and silicate. From Figure 17 it is clear that debris is thrown out in all directions. Whether it is retained in the Solar System or how far this material moves away from the Sun depends on its intrinsic energy. This was found separately for core, mantle and ice material for those SPH particles outside the spheres of influence of the two cores. Figure 22 shows the mass in Earth units per unit intrinsic energy, expressed in GJ/kg.

More than one-half of the core material is retained and a somewhat smaller fraction of mantle material. However the majority of the ice is expelled from the system. The further from the centre the silicate material was within the planets the higher was its volatile content and the less constrained its motion, so that it would have travelled out further from the collision. From observations, as judged by their reflection spectral features, the C-type asteroids, containing the most volatile material, tend to be further from the Sun than S and M-type asteroids, associated with silicate and metal compositions – which is consistent with the model suggested here.

The eccentricities and inclinations of the retained material show some interesting trends, as seen in Figure 23. Nearly all the iron and mantle debris orbits are prograde but a considerable proportion of the ice orbits are retrograde.

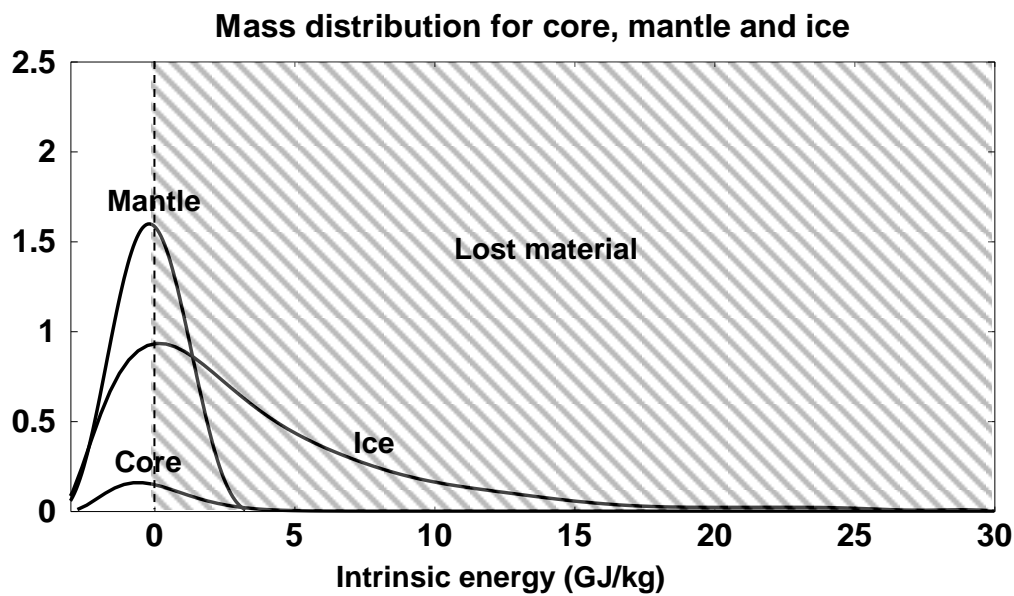

Figure 22  The distribution of dispersed material from the collision. The graphs show the distribution of mass in Earth units per unit intrinsic energy (GJ/kg) (Woolfson, 2013b).

Asteroid bodies, considered as debris, would have interacted with each other and with planets. Some would have attained safe orbits, such as those in the Asteroid Belt or in the region between Saturn and Uranus now occupied by Chiron and probably other bodies. The total mass of the surviving asteroids, about 4% of that of the Moon, is small since most debris would have been swept up by the major planets once they had settled into their final orbits.

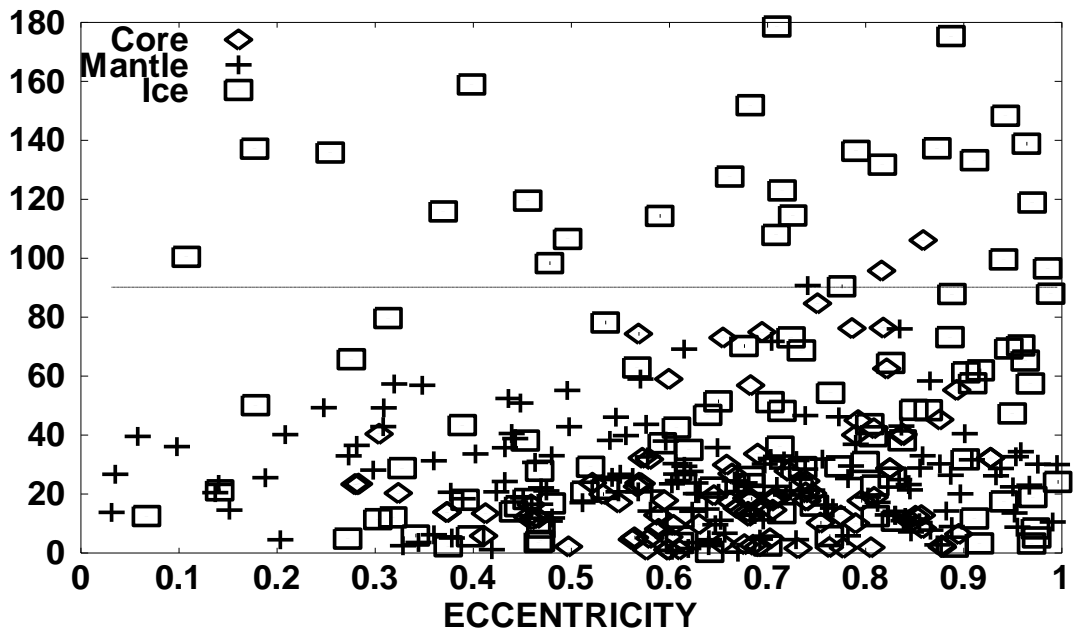

Figure 22  The eccentricities and inclinations of the retained material

Observed comets are associated with two regions of the Solar System – the Kuiper Belt, just beyond the orbit of Neptune, and the Oort Cloud, tens of thousands of au from the Sun. It has been argued that Oort-Cloud comets have an external origin because they have D/H ratios 20 times the cosmic value but the high D/H values in the ice of the colliding planets weakens that argument. Bailey (1983) proposed that there is an inner reservoir of comets, between the Oort Cloud and the Kuiper Belt, which is drawn outwards to replenish the Oort Cloud when it is depleted by a very severe disturbance, such as when a star passes near or through it or when the Solar System passes through a Giant Molecular Cloud. The results in Figure 22 are consistent with the Bailey model since the intrinsic energies of both the mantle and ice debris give a continuous distribution of debris stretching from the inner solar system to the region of the Oort Cloud. If the continuity of the distribution has persisted throughout the lifetime of the Solar System then the observed parts of the Kuiper Belt form the inner boundary of this distribution. Short-period comets, perturbed by Neptune, come from the Kuiper Belt. The so-called *new comets* are those from the outer part of the distribution, the Oort Cloud, which are perturbed by sources exterior to the Solar System. Since there are no major sources of perturbation for the great majority of comets at the centre of the distribution their presence is not detected.

A characteristic of orbits is that if the only forces at play are the gravitational forces of the two principal bodies then they repeatedly go through the same regions of space. Hence, once the planetary orbits had stabilized, a fragment beginning in the terrestrial region will repeatedly pass through the region occupied by the planets and inevitably it will eventually collide with one of them or be thrown out of the Solar System. Only if the perihelion is beyond Neptune will it have a permanent, or at least long-term, existence.

Some material, with more volatile content that went further out, could have interacted with major planets in evolving orbits at hundreds of au from the Sun and be swung into orbits well outside the present planetary region. With aphelion = $Q$ and perihelion = $q$, Figure 24 shows an interaction of a comet with $(Q, q)) = (110$ au$, 0.5$ au$)$ with a Jupiter-mass planet with $(Q, q) = (100$ au$, 10,$ au$)$ with closest approach $1.84 \times 10^6$ km. After the interaction, for the comet $(Q, q) = (109.9$ au$, 42.1$ au$)$, which would place the body well within the Kuiper Belt. This process is most efficient if the interacting bodies have similar aphelia so it is unlikely that many fragments will be affected in this way.

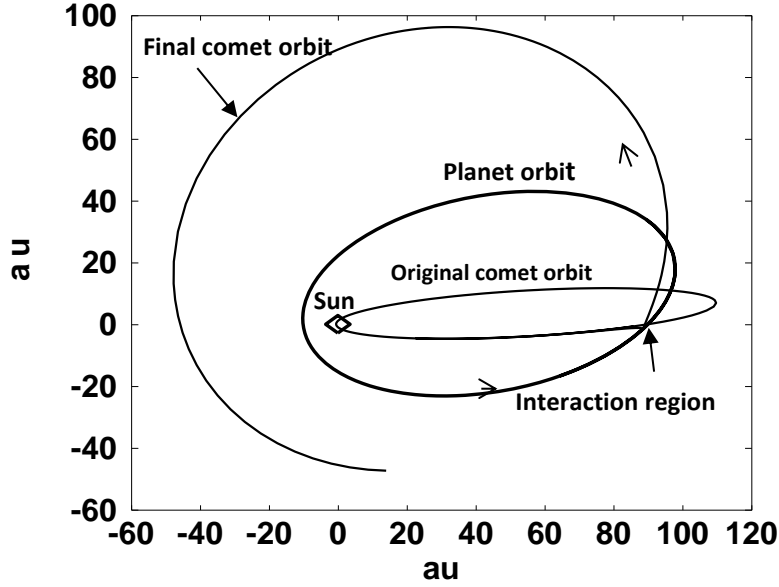

Figure 24  Debris interacting with a planet in an evolving orbit is perturbed into an orbit with a small perihelion to a much larger one.

The the evolution of planetary orbits in a resisting medium, either by the Type-I or Type-II processes, are due to the planetary mass affecting the medium with the resistance coming from the reaction on the planet. Bodies as small as comets do not have sufficient mass greatly to influence the distribution of the medium and the resistance is primarily due to the ram pressure they experience because of the impact of the medium on the comet. For a spherical comet the force experienced will be

$$\mathbf{F} = \pi \rho a^2 \mathbf{V} \qquad (16)$$

where $\rho$ is the local density of the medium, $a$ the radius of the comet and $\mathbf{V}$ the velocity of the medium relative to the comet. The effect of such a force has been found for a comet of mass $7 \times 10^{12}$ kg, of density 500 kg m$^{-3}$ (published estimates are between 100 and 1,000 kg m$^{-3}$), with original perihelion 0.5 au and a range of original semi-major axes. The medium had a total mass of 40 $M_J$ with an annular distribution of density, similar to that seen in Figure. 5, given by

$$\rho(r,z) = C \exp\left(-\frac{(r-d)^2}{2\sigma_r^2}\right) \exp\left(-\frac{hr^2}{s^2}\frac{z^2}{\sigma_z^2}\right) \qquad (17)$$

where $d = 100$ au, $\sigma_r = 30$ au, $h = 10$, $s = 20$ au, $\sigma_s = 30$ au and $r$ and $z$ (distance from the mean plane) are expressed in au. The constant $C$ is determined from the total mass of the disk.  The results are shown in Figure 25

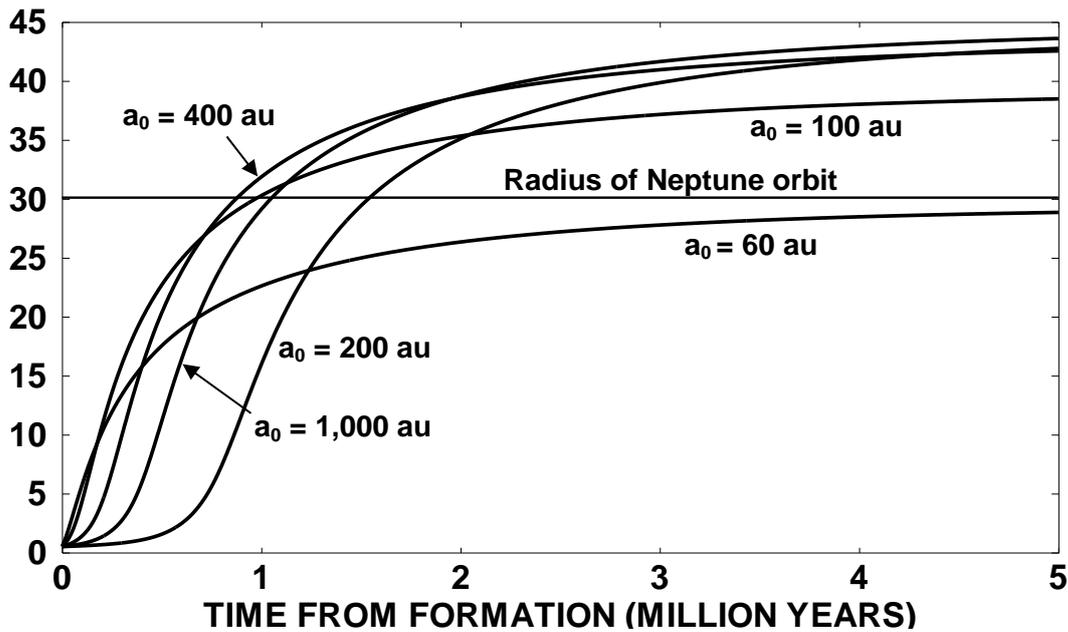

Figure 25  The changes of the perihelion with time for comets with original perihelia 0.5 au and various initial semi-major axes, $a_0$ (Woolfson, 2013b).

It will be seen from the figure that all orbits with original semi-major axes greater than about 60 au will end up with perihelia beyond the orbit of Neptune and within the Kuiper Belt region. Their aphelia stretch out to several hundred and even thousands of au and represent the inner cloud of comets postulated by Bailey as a reservoir for the replenishment of the Oort Cloud.

## 14  Dwarf planets

We have already identified the Moon, Mars, Mercury and Triton as onetime satellites of the colliding planets but there would have been many more. There is now a class of bodies known as *dwarf planets*, six in number – Ceres, Pluto, Eris, Makemake, Haumea and V774104, the last having been discovered at the end of 2015. They all have masses within the range of solar-system satellites and we identify them as ex-satellites of the colliding planets.

The orbits of bodies of satellite mass are modified by the Type 1 migration process in which the body maintains contact with the medium, but the effectiveness of the process depends on the mass of the satellite as well as the density of the medium. Since orbital evolution is slow for smaller-mass bodies it is unlikely that orbital evolution will progress to the round-off stage within the lifetime of the circumstellar disk. From Figure 7 it is seen that the perihelion increases up to the time that round-off occurs. Satellites with orbits that evolved so that the perihelion was within the Kuiper Belt would survive while others would eventually be swept up by a major planet or expelled from the Solar System. The exception is Ceres that was left in an orbit within the asteroid belt.

There could have been many other ex-satellites, possibly some more massive than the present dwarf planets. Some of these might have escaped from the Solar System directly as a result of the collision and others could have been absorbed by major planets. Others, yet to be discovered, might be in outer regions of the Kuiper Belt. Another possibility is that one or more of them may have reached the Oort Cloud. It has sometimes been suggested that the tendency of some new comets to come from similar directions with similar orbital parameters may be due to the presence of major perturbing bodies within the Oort Cloud itself.

If dwarf planets were redefined as being ex-satellites of the originally-formed planets, large enough for their self-gravity to mould them into hydrostatic equilibrium and in heliocentric orbits then, according to the model suggested here, at present there would be eight – Ceres, Pluto, Eris, Makemake, Haumea, V774104, Mars and Mercury.

## 15 The ice giants

The ice giants are distinctive in having low masses, a high proportion of volatile materials that, when cold, form ices, high spin-axis tilts and atmospheric D/H ratios much higher than both the galactic average and that of the gas-giant planets. It has often been postulated that a large body colliding with Uranus produced its present extreme spin-axis tilt of $97.8^o$ but to achieve such a a change of angular momentum would have required a large body striking Uranus at a very high speed. If we assume that Uranus has been unchanged in structure and composition since its formation and that its spin axis initially had zero tilt then the magnitude of the imparted angular momentum required to change it to its present configuration is of order $10^{38}$ kg m$^2$ s$^{-1}$. If the body struck Uranus tangentially then the momentum of the body, relative to Uranus would have been about $4 \times 10^{30}$ kg m s$^{-1}$, which could be provided by a body of mass 7 $M_\oplus$ with speed 100 km s$^{-1}$.

A clue as to the history of Uranus and Neptune is given by the D/H values in their atmospheres. The values for the two pairs of giant planets are;

Jupiter, $2.25 \pm 0.35 \times 10^{-5}$ and Saturn, $1.70^{+0.75}_{-0.45} \times 10^{-5}$ (Lellouch et al., 2001)
Uranus, $4.4 \pm 0.4 \times 10^{-5}$ and Neptune, $4.1 \pm 0.4 \times 10^{-5}$ (Feuchtgruber et al., 2013)

As an illustration we make the the assumption that the star forming cloud had the overall D/H universal ratio of $2 \times 10^{-5}$, that 75% by mass of the gas in the cloud was hydrogen and that hydrogen in ice, with a D/H ratio of 0.015, formed 0.1% of the mass of the cloud. On this basis the D/H ratio in the gas was $6.7 \times 10^{-6}$. The ratio of hydrogen in gas form to that contained in ices is 750:1 and any planet with a lesser ratio will have a final D/H ratio greater than the universal value once the deuterium has diffused from the ice into the atmosphere.

For the NT the spin-axis tilts can be explained either by the way in which the cores were produced by planetesimal accretion or by impacts. Again, since the mass of the atmosphere is not wholly dependent on the core mass but also on the density and duration of disk material in its vicinity, the ratio of atmospheric hydrogen to hydrogen in ices could have been much lower than 750:1, so explaining the high D/H ratio.

For the CT the original spin-axis tilt should not have been large. Condensations in the filament can be either above or below the Jeans critical mass. If below the critical mass they can still form a stable condensation from central material despite losing a great deal of outer material and if above critical mass they might still lose some outer material before producing a stable condensation (Ruskol, 1960; Woolfson, 1964).

Here we describe a scenario, consistent with the CT, which both explains the large axial tilts of the ice giants and also influences the D/H ratio. It proposes that the ice giants originally had more extensive atmospheres that were partially stripped off in a close oblique collision with a much more massive planet soon after their formation so that the ices within them still retained the original large D/H ratio. This collision also provided the change of angular momentum to tilt the spin axes,. To illustrate this we show an SPH simulation of a collision between a proto-ice-giant (PIG) and a much more massive major planet, which we take as Bellona. Point-mass models for the planets were formulated, as described by Woolfson (2007), based on a planet in four distinct layers. For recently-formed planets segregation by density would not be complete so the four layers had compositions as specified below.

    core, consisting of 40% iron with the remainder silicate,
    mantle, consisting of 85% silicate with the remainder iron,
    silicate+ice layer with 10% of the mass (~ 25% by volume) as ice,
    atmosphere, consisting of hydrogen and helium.

As described previously, the masses of the core, mantle and ice layer were specified and the Tillotson (1962) equations of state or iron, granite (representing silicate materials) and water (representing ices) were used to integrate outwards starting with a specified central temperature and density. Previously published models of Jupiter and Saturn (Stevenson and Salpeter, 1976) were used to choose a best model for an atmosphere of the form

$$p = \frac{\rho k T}{\mu}(1 + c\rho), \qquad (12)$$

which gave $c = 0.08$. The PIG had mass $54.58 M_\oplus$ and radius $5.049 \times 10^4$ km. The compositions, by mass, of the inner three layers are given in Table 5. The density of points, 3,319 in total for each body, is highest at the centre of each body, the region of greatest interest, and falls off towards the boundary. This number of points representing each body is small by modern standards but it was sufficient to give a faithful representation of the collision event and enabled a precise pairs-representation of gravity. An initial configuration of the two planets is shown in Figure 26; the motion of the PIG along the $x$-direction is displaced relative to Bellona along the $y$ direction by a distance $10^5$ km. Both bodies initially have spin axes in the $z$ direction with a spin period 10 hours, similar to the periods of Jupiter and Saturn.

The progress of the oblique collision can be followed in Figure 27, which shows the arrangement of the two bodies at four times relative to zero time for Figure 26. The PIG is considerably disrupted and distorted, as is seen in frames (a) and (b), but quickly reassembles itself into a spherical form.

The mass and radius of the residual ice giant (RIG), as seen in Figure 27(d) are given in Table 6. The RIG retains all the core, mantle and ice material of the PIG so the final mass of the atmosphere is $10.18 M_\oplus$, a fraction 0.222 that of the original atmospheric mass of $45.88 M_\oplus$. The mass of hydrogen in the final atmosphere is $0.75 \times 10.18 M_\oplus = 7.635 M_\oplus$. We now have to estimate how much hydrogen is contained in the $2.5 M_\oplus$ of material described as ice, which is really ice-impregnated silicate material. Taking 0.1% as hydrogen, the estimated mass of hydrogen in the ice layer is $0.025 M_\oplus$. Thus for the RIG the ratio of hydrogen in the atmosphere to that in ice is 305 giving the value of D/H shown in the table.

Table 5  The inner composition of the colliding planets

| Planet | Core($M_⊕$) | Mantle ($M_⊕$) | Ice ($M_⊕$) |
|---|---|---|---|
| Bellona | 3.00 | 12.00 | 6.00 |
| Proto-ice-giant | 1.25 | 5.00 | 2.50 |

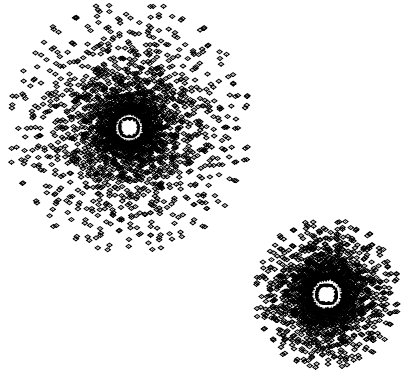

Figure 26  The initial configuration for the grazing collision

The characteristics of the RIG are clearly those that we associate with an ice giant. The large tilt is brought about by the direct transfer of angular momentum by the impact plus the huge tidal effect Bellona exerts on the PIG, especially during the close approach of the two bodies. By varying the parameters of the model other outcomes are possible, with more or less mass and radius and with a greater or lesser inclination.

This model is purely speculative and illustrates a possible way in which the axial tilt and D/H ratios may be linked in a single event.  However, as indicated previously, the D/H ratio may just come about just as a consequence of the formation process and the tilt may result in an event that left the ice giant structurally intact. However, it does seem likely that any event so violent that it would cause a massive change in axial tilt would also affect the structure of an ice giant.

Table 6 Characteristics of Uranus, Neptune and the RIG

|  | Uranus | Neptune | RIG |
|---|---|---|---|
| Mass ($M_⊕$) | 14.54 | 17.15 | 18.93 |
| Radius ($10^4$ km) | 2.556 | 2.476 | 2.543 |
| Mean density ($10^3$ kg m$^{-3}$) | 1.27 | 1.64 | 1.64 |
| Spin-axis tilt (°) | 97.8 | 28.3 | 68.7 |
| D/H ($10^{-5}$) | 4.4 | 4.1 | 4.91 |
| Moment-of-inertia factor | 0.225 | unknown | 0.220 |
| Spin period (hours) | 17.24 | 16.11 | 17 |

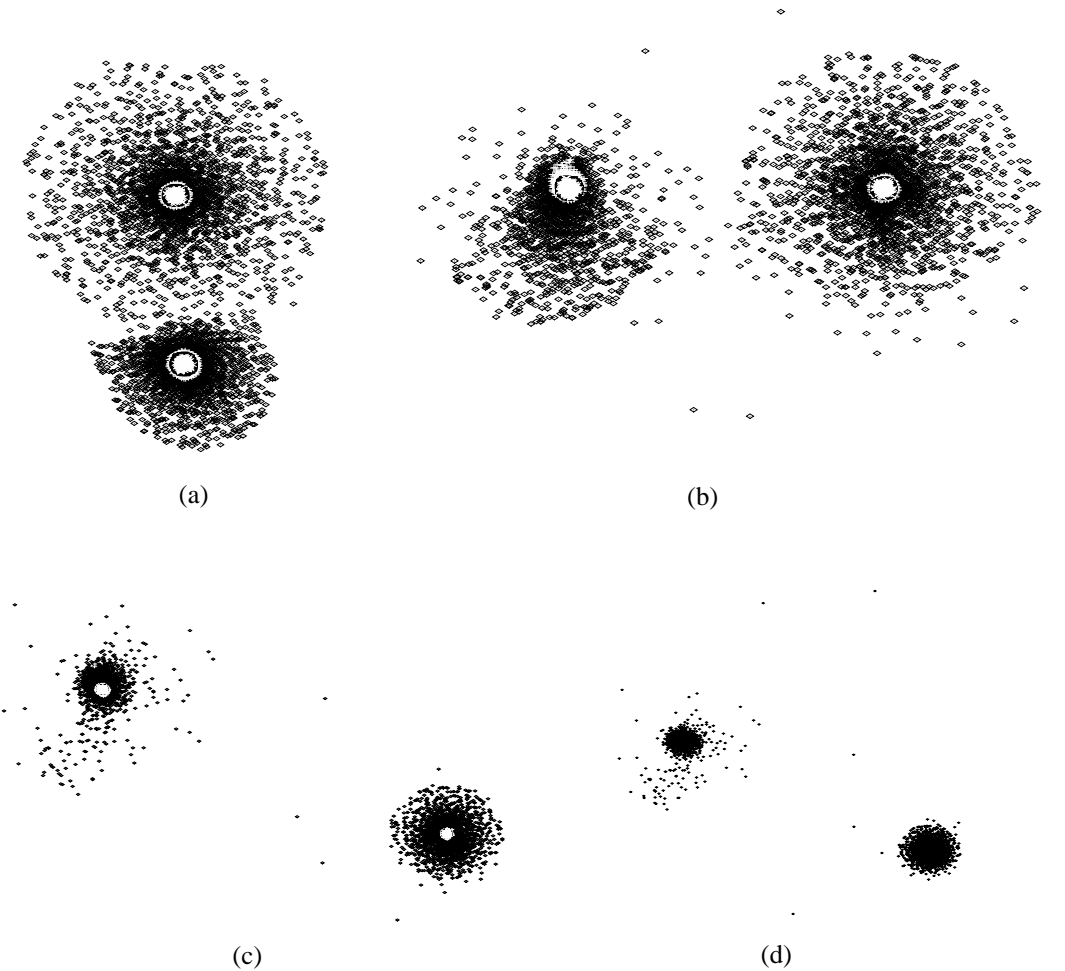

Figure 27  The progress of the planetary oblique collision after (a) 0.56 hr, (b) 1.33 hr, (c) 3.31 hr, (d) 4.18 hr. The scales vary but the Bellona condensation remains at an approximately constant size.

The axial tilt of Saturn, $27^o$, is similar to that of Neptune, $29^o$. If it was involved in an axial-tilt-modifying interaction and *gained* atmospheric gas in the process then this would explain its somewhat low value of D/H.

## 16  Isotopic anomalies in meteorites

Many meteorites show features suggesting the onetime presence of silicate vapour, which would have come naturally from the planetary collision. Chondritic meteorites contain solidified silicate droplets, *chondrules*, that condensed from a vapour and minerals condensed in a sequence controlled by their condensation temperatures as the vapour cooled (Larimer, 1967).

The isotopic compositions of some meteorites are very different from those of terrestrial material; where this occurs the difference is denoted as an *isotopic anomaly* These are described here for the elements carbon, nitrogen, oxygen and neon, but there are many others.

In terrestrial carbon the ratio of the two stable isotopes $^{12}_{6}C:^{13}_{6}C$ is 89.9:1. In mineral grains of silicon carbide, SiC, contained in some chondritic meteorites the ratio is much less, down to 20:1. Since there is more of the heavier isotope this anomaly is known as *heavy carbon*. The usual explanation is that it is due to grains containing heavy carbon coming into the Solar

System from distant carbon stars and that the observed anomalies in silicon carbide grains originated in six or more carbon stars, each with carbon of different heaviness..

Silicon carbide also contains nitrogen trapped in grain interstices. The ratio of the two stable nitrogen isotopes on Earth, $^{14}_{7}N:^{15}_{7}N$, is 270:1. Most SiC-derived nitrogen is *light nitrogen* with ratios up to 2,000:1 but, rarely, *heavy nitrogen* occurs with ratios down to 50:1.

The three stable isotopes of oxygen have ratios in terrestrial samples given by

$$^{16}_{8}O : ^{17}_{8}O : ^{18}_{8}O = 0.9527 : 0.0071 : 0.0401$$

This mixture, known as SMOW (Standard Mean Ocean Water), also occurs in Moon rocks. Two kinds of meteorite contain oxygen isotopic ratios that cannot be explained by processing terrestrial oxygen. These are ordinary chondrites and carbonaceous chondrites, the latter being stony meteorites that contain volatile material. The oxygen anomalies can be explained as the addition of different amounts of pure, or nearly pure, $^{16}_{8}O$ to SMOW. The usual explanation again involves grains drifting across interstellar space and entering the Solar System. The $^{16}_{8}O$ is assumed to have been produced by the action of alpha particles on $^{12}_{6}C$ in distant stars. Subsequently normal solar-system oxygen infiltrated the grains displacing most, but not all, of the $^{16}_{8}O$ so giving a final content that has a surplus of that oxygen isotope.

The final example of an isotopic anomaly is for neon. The three stable isotopes have terrestrial ratios

$$^{20}_{10}Ne : ^{21}_{10}Ne : ^{22}_{10}Ne = 0.905 : 0.003 : 0.092$$

Neon atoms are trapped in atomic-size cavities, but can be released by heating the meteorite. If neon is found in meteorites then they could not have been substantially heated after the neon was incorporated. The same is true for other gases trapped in the interstices of meteorite grains.

Isotopic compositions of neon in meteorites are very variable and it has been deduced that they come from admixtures of three neon sources with different compositions. However, some meteorites contain pure, or almost pure, $^{22}_{10}Ne$ that is a 9.2% component of terrestrial neon. This anomalous neon is called *neon-E*. Since $^{22}_{10}Ne$ cannot be separated from a mixture of isotopes some other explanation is required. Sodium has one stable isotope, $^{23}_{11}Na$, but there is a radioactive isotope, $^{22}_{11}Na$, which decays into $^{22}_{10}Ne$. One suggested scenario is that just before solar-system formation a nearby supernova produced $^{22}_{11}Na$ that was incorporated with stable sodium, in minerals. The $^{22}_{11}Na$ decayed and the resultant $^{22}_{10}Ne$ was trapped within the mineral grains. A problem with that scenario is that $^{22}_{11}Na$ has a short half-life, 2.6 years, so that, after production in a supernova, it has to be incorporated into a *cool* solid body within a period of 10-20 years. This puts a very tight constraint on the timing of the supernova.

Holden and Woolfson (1995) examined the effect of subjecting a mixture of iron, silicates and ices, such as will occur in protoplanets, with a D/H ratio that of Venus (0.016), to a triggering temperature of $3 \times 10^6$ K. The rates, given by Fowler, Caughlan and Zimmerman (1967, 1975), for 548 nuclear reactions were used and 40 decay processes were involved in their calculation. All possible cooling factors were included; iron, which took no part in reactions was a coolant and ionization of material, implemented by a solution of the Saha equations (Zel'dovich and Raiser, 1966), greatly increased the number of particles present to share the generated energy. The outcome was a nuclear explosion, the products of which explained a number of important light-atom isotopic anomalies. The final temperature was well in excess of $5 \times 10^8$ K.

Another calculation by Woolfson (2011b), using a lesser D/H ratio of 0.01, gave similar results, explaining all anomalies for carbon, nitrogen, oxygen, neon, magnesium, aluminium and silicon. This single explanation replaced a number of *ad hoc* explanations for individual anomalies.

A large quantity of $^{13}_{6}C$, and radioactive $^{13}_{7}N$ that decays to $^{13}_{6}C$ with a half-life of 9.97 minutes, was produced, which explained the full range of heavy-carbon observations. Figure 28 shows the concentrations of $^{12}_{6}C$ and $^{13}_{6}C$ during the explosion; the $^{13}_{6}C$ concentration includes the contribution of $^{13}_{7}N$. When the temperature exceeds $3 \times 10^8$ K there is a sharp increase in the amount of $^{13}_{6}C$.

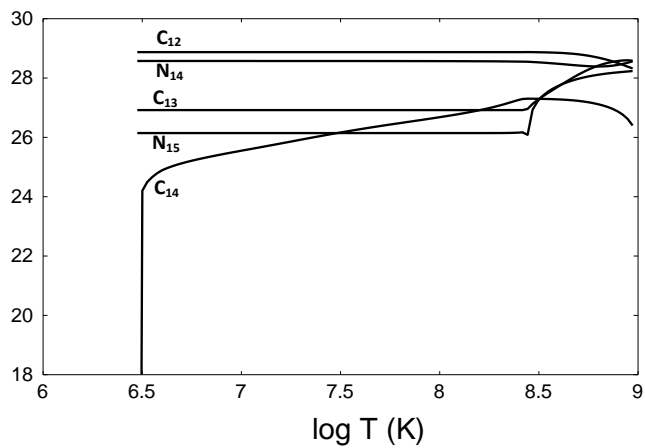

Figure 28 The concentrations of isotopes of carbon and nitrogen as the temperature within the nuclear-reaction region increases (Woolfson, 2013b)

There is a small reduction in the amount of $^{14}_{7}N$, including the contribution of $^{14}_{8}O$ that quickly decays to $^{14}_{7}N$, as the explosion progresses, although it picks up at very high temperatures (Figure 28). In the long term the concentration of $^{14}_{7}N$ is augmented by $^{14}_{6}C$ decay with a half-life of 5,739 years. Starting with $^{14}_{7}N$ and $^{15}_{7}N$ trapped inside the grains once they were cool enough to retain the gas, during the next few tens of thousands of years $^{14}_{6}C$, which had been part of the silicon-carbide host, decayed, with the additional $^{14}_{7}N$ boosting the amount of that isotope, so giving light nitrogen. If the original amount of the two stable nitrogen isotopes was not very great then the $^{14}_{6}C$ contribution can make a large proportional change to the $^{14}_{7}N$ concentration.

Towards the end of the explosion there is an almost 100-fold increase in the amount of $^{15}_{7}N$ present, produced by reactions involving heavier elements. These elements may not be uniformly distributed so that $^{15}_{7}N$-rich pockets can form. Silicon-carbide grains within these pockets could take up $^{15}_{7}N$ - rich nitrogen and, even after the decay of $^{14}_{6}C$, the result can be the occasional occurrence of heavy nitrogen.

Figure 29 shows the variation of stable oxygen isotopes during the explosion. The $^{17}_{8}O$ and $^{18}_{8}O$ concentrations include contributions from the fluorine radioactive isotopes, $^{17}_{9}F$ and $^{18}_{9}F$, which quickly decay to $^{17}_{8}O$ and $^{18}_{8}O$, respectively. Above about $6 \times 10^8$ K, the concentrations of $^{17}_{8}O$ and $^{18}_{8}O$ greatly diminish leaving virtually pure $^{16}_{8}O$ that, mixed with SMOW in various proportions, gives the oxygen isotopic anomaly.

A sufficient quantity of $^{22}_{11}Na$ was produced in the explosion to explain the production of neon-E. The scale of a planetary collision is small by astronomical standards and the formation of cool grains, which would retain $^{22}_{10}Ne$, would take place within hours or days (Woolfson, 2011c).

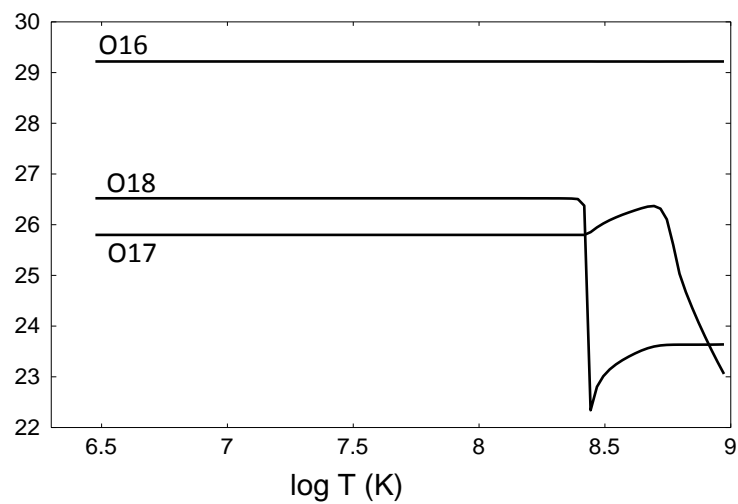

Figure 29 The variation of the isotopes of oxygen (including radioactive fluorine) with temperature (Woolfson, 2013b)

## 17 Summary and conclusions

This review has covered a great deal of ground, from the formation of a star-forming cloud from ISM material to details of solar-system evolution. To summarize all this material and bring it into perspective the steps have been:

(1)  Describing the action of a supernova in precipitating a star-forming cloud.
(2)  Noting that observations indicate that cloud collapse leads to a dense embedded cluster within which protostars and colliding turbulent gas streams co-exist with YSOs and main-sequence stars.
(3)  Simulating the CT process with both a protostar, and a dense region produced by colliding gas streams, to produce a gravitationally-unstable dense filament within which protoplanetary blobs form. These initially move in extensive, highly-eccentric orbits
(4)  Showing that some protostar material forms a circumstellar disk within which orbits evolve, usually by round-off and decay. In some circumstances the final outcome is high-eccentricity orbits. Interactions between pairs of planets during orbital evolution can give commensurate orbits. All types of exoplanet orbit can be reproduced.

(5) Describing how satellites are produced on a short timescale by an accretion process within dense disks left behind by collapsing protoplanets.

(6) Explaining how absorption of protostar material by the star, either directly or by the inward drift of the circumstellar medium, moves the initially randomly-oriented stellar spin axis closer to the normal to the exoplanet orbits, hence reducing spin-orbit misalignment. This gives some retrograde orbits but a strong bias towards low-inclination direct orbits.

(7) Simulating the motion of a protostar to show that the number of CT interactions within a dense embedded cluster can explain the currently estimated proportion of stars with planets.

(8) Due to grain-surface chemistry within star-forming clouds early planets contain shells of icy materials with high D/H ratios, surrounding a central iron-silicate core. A collision between two erstwhile giant planets – Enyo and Bellona – triggered D-D nuclear reactions, the energy from which fragmented the planets. Two residual cores formed the Earth and Venus.

(9) It is proposed that the Moon is a normal ex-satellite of Bellona retained by the Earth core. Hemispherical asymmetry of the Moon is explained as due to abrasion of the hemisphere facing the collision due to high-speed debris.

(10) Mars and Mercury are explained as two large ex-satellites released into heliocentric orbits. The hemispherical asymmetry of Mars and high density of Mercury are due to abrasion by collision debris. The comparatively large eccentricities of their orbits are due to incomplete round-off, since this is slow for small-mass bodies.

(11) Asteroids and comets are derived from collision debris. Their distributions depend mainly on their initial intrinsic energy and interactions with the circumsolar disk. Apart from the small numbers in the Asteroid Belt, their long-term survival depended on their orbits evolving to give a perihelion outside the orbit of Neptune.

(12) Dwarf planets are those ex-satellites of the colliding planets whose heliocentric orbits evolved to give a perihelion outside the orbit of Neptune and within the Kuiper Belt.

(13) The relatively high D/H ratios and axial tilts of ice giants can be explained as the result of original bodies with more extensive atmospheres having close interactions with massive planets.

(14) Light-atom isotopic anomalies are explained as the outcome of the nuclear reactions that occurred in the planetary collision. This model replaces many ad-hoc explanations for individual anomalies.

Figure 30 is a visual summary that shows the interconnectivity of the processes that occur. All the physical processes involved are well understood – for example, tidal interactions, collisions and nuclear reactions – and occur in many astronomical contexts. There are no improbable, exotic or poorly-understood processes that occur at any stage.

The planetary collision that initiates the evolution of the Solar System to its present state occurred because the precession of planetary orbits gave intersecting orbits from time-to-time. The plausibility of such a collision event is supported both by analysis of orbital precession and by an observation of such an event having taken place in the vicinity of a young star within the last few thousand years. It is interesting to note that the circumsolar disk had two roles to play, both to give a resisting medium and, through its gravitational influence, to give precession of the planetary orbits. That a single event leads to the explanation of so many features of the Solar System in straightforward ways increases the plausibility of the postulate that a planetary collision might have occurred. It also fits well with the Occam's-razor principle that a plausible theory should explain as much as possible with a minimum number of assumptions.

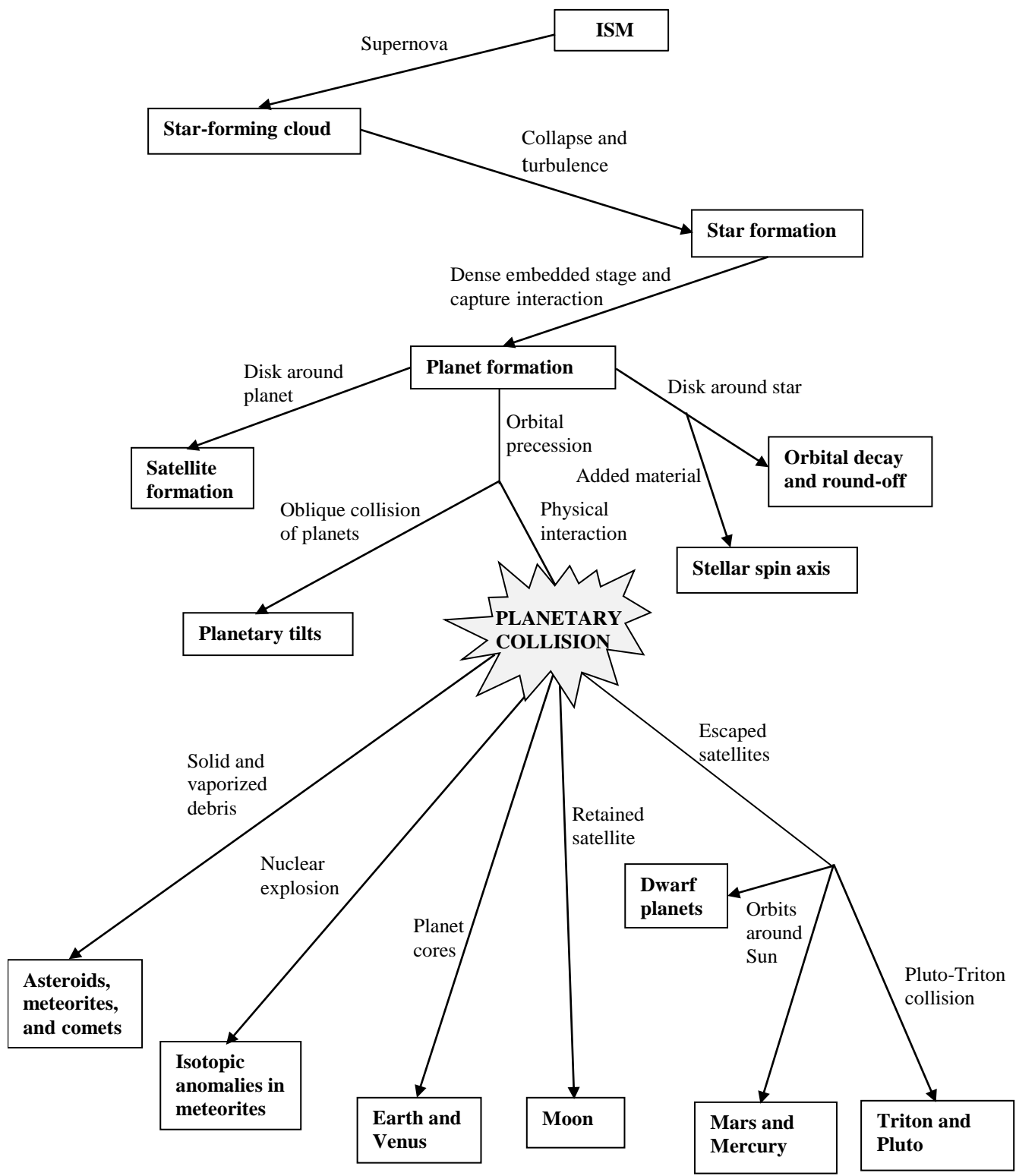

Figure 30 Consequences of the Capture Theory and a planetary collision

# References

Anderson D. R. et al. (2010) *Astrophys. J.*, **709**, 159-167

Armitage, P.J. and Clarke, C.J. (1996) *Mon. Not. R. Astr. Soc.*, **280**, 458-468

Bailey, M.E. (1983) *Mon. Not. R. Astr. Soc.*, **204**, 603-633

Bailey, V et al**. (**2014) *Astrophys. J. L.,* **780** L4.

Benz, W., Slattery, W.L. and Cameron, A.G.W. (1986) *Icarus*, **66**, 515-535

Bonnell, I.A, Bate, M.R. and Vine, S.G. (2003) *Mon. Not. R. Astr. Soc.*, **349**, 413-418.

Bonnell, I.A., Bate, M.R. and Zinnecker, R. (2005) *Proc. I,A.U. Sympoaium*, No.227, eds. R. Cesaroni, M. Felli, E, Churchwell and C,M. Walmsley

Borucki, N.J. et al. (2011) *Astrophys. J.*, **736**, 19-40

Cole, G.A.H. and Woolfson, M.M. (2013) *Planetary Science*: *The Science of Planets around Stars*, 2$^{nd}$ Ed. p. 501, CRC Press, Boca Raton

Cook, A.H. (1977) *Celestial masers*, Cambridge University Press: Cambridge

D'Angelo, G. and Lubow, S.H. (2010) *Astrophys. J.* **724**, 730-747.

Dormand, J.R. and Woolfson, M.M. (1977) *Mon. Not. R. Astr. Soc.*, **180**, 243-279

Fabrycky, D.C. and Tremaine, S., (2007) *Astrophys. J. 669*, 1298-1315

Feuchtgruber, H. et al., 2013, arXiv:1301.5781[astro-ph.Ep]

Fowler, W.A., Caughlan, G.R. and Zimmerman, B. (1967) *Ann. Rev. Astron. Ap.*, **5**, 525-576.

Fowler, W.A., Caughlan, G.R. and Zimmerman, B. (1975) *Ann. Rev. Astron. Ap.*, **13**, 69-112.

Gaidos, E.J. (1995) *Icarus*, **114**, 258-268

Golanski, Y. and Woolfson, M.M. (2001) *Mon. Not. R. Astr. Soc.*, **320**, 1-11.

Goldreich, P. and Ward, W.R. (1973) *Astrophys. J.* **183**, 24-26.

Hayashi, C. (1966) *Ann. Rev. Astron. Astrophys.,* **4**, 171-192

Head, J.W. (1976) *Reviews of Geophysics and Space Physics*, **14**, 265-300

Heller, R., htpp//ooo.aip.de/People/rheller/content/main_spinorbit.html, 20$^{th}$ June, 2013.

Holden, P. and Woolfson, M.M. (1995) *Earth, Moon and Planets*, **69**, 201-236

Indulekha, H. (2013) *J. Astrophys. Astron.*, **34**, 207-215

Jacobson, S.A. et al. (2014) *Nature*, **508**, 84-87

Jeans, J. H. (1902) *Phil. Trans. R. Soc. Lond.*, A**199**, 1-53

Jeans, J.H. (1917) *Mon. Not. R. Astr. Soc.*, **77**, 186-199

Kozai, Y. (1962) *Astrophys. J.* **67**, 591-598

Kroupa, P. (2001) *Mon. Not. R. Astr. Soc.*, **323**, 231-246

Lamy, P.L. and Burns, J.A. (1972) *A. J. Phys.* **40**,441-444

Laplace, P.S. de (1796) *Exposition du Système du Monde.* Imprimerie Cercle-Social, Paris

Larimer, J.W. (1967) *Geochimica et Cosmochimica Acta,*, **31**, 1215-1238

Lellouch, E. et al., 2001, *A&A*, **670**, 610-622.

Loinard, L. et al. (2001) *Astrophys. J.*, **552**, L163-166

Loinard, L. et al. (2002) *Planet. Space Science*, **50**, 1205- 1213

Lubow, S.H. and Ida,S (2011) *Planet Migration*, Ed. S. Seager, pp. 347-371, University of Arizona Press, Tucson, AZ

Lucas, P. and Roche, P. (2000) *Mon. Not. R. Astr. Soc.*, **314**, 858-864

Lynden-Bell, D. and Pringle, J.E. (1974) *Mon. Not. R. Astr. Soc.*, **168**, 603-637

McConnell, A.J. and Woolfson, M.M. (1983) *Mon. Not. R. Astr. Soc.*, **204**, 1221-1240

McCord, T.B. (1966) *Astron, J.*, **71**, 585-590

McLaughlin, D.B. (1924) *Astrophys. J.*, **60**, 22-31

McCaughrean, M.J. and Stauffer, J. (1994) *AJ*, **108**, 1382-1397

Melita, M.D. and Woolfson, M.M. (1996) *Mon. Not. R. Astr. Soc.*, **280**, 854-862

Mullis, A.M. (1993) *Geophys. J. Int.*, **114***,* 196-208

Nutzman, P.A., Fabrycky, D.C. & Fortney, J.J. (2011) *Astrophys. .J. L.*, **740**, 10.
Observatoire de Paris (2004) http://obspm.fr/how-did-the-planet -in-the-gamma-cephei-binary.html
Oxley, S. and Woolfson, M.M. (2003) *Mon. Not. R. Astr. Soc.*, **343**, 900-912.
Oxley, S. and Woolfson, M.M. (2004) *Mon. Not. R. Astr. Soc.*, **348**, 1135-1149
Parise, B. et al. (2002), *Astron. & Astrophys.*, **393**, L49-53
Protszkov, E-M, Adama, F.C., Hartmann, L.W. and Tobin, J.J. (2009) *Astrophys. J.,* **697**, 1020-1032
Rogers, T.M., Lin, D.N.C. and Lau, H.H.B.,(2012) *Astrophys. J. Let.*, **754**, L6-L10
Rossiter, R.A.. (1924) *Astrophys. J.*, **60**,15-21
Roueff, E. et al. (2000) *Astron. Astrophys.*,**354,** L63-66
Ruskol, E.L. (1960) *Sov. Astron. AJ*, **4**, 657-668
Safronov, V.S. (1972) *Evolution of the Protoplanetary Cloud and Formation of the Earth and Planets* (Israel Program for Scientific Translation, Jerusalem).
Seaton, M. (1955) *Ann. Astrophys.* **18**, 188-205.
Stevenson, D.J. and Salpeter, E.E. (1976) *Jupiter*, Ed.T.Gehrels, pp. 85-112, University of Arizona Press, Tucson, AZ
Sumi, T. et al. (2011) *Nature*, **473**, 349-352
Tillotson, J.H. (1962), Tec. Rep. General Atomic Report G-3216.
Weidenschilling, S.J., Dunn, B. and Meakin, P. (1989) *The Formation and Evolution of Planetary Systems*, eds. H.A. Weaver and L. Danley (Cambridge University Press, Cambridge). pp. 131-150.
Williams, I.P. and Cremin, A.W. (1969) *Mon. Not. R. Astr. Soc.*, **144**, 359-373.
Wood, K., Wolff, M.J., Bjorkman, J.E. and Whitney, B. (2001) *Astropys. J.*, **564**, 889-892
Woolfson, M.M. 1964) *Proc. R. Soc.* A**282,** 485-507
Woolfson, M.M. (1979) *Phil. Trans. R. Soc. Lond.*, A**291**, 219-252
Woolfson, M.M. (1999) *Mon. Not. R. Astr. Soc.*, **304**, 195-198
Woolfson, M.M. (2003) *Mon. Not. R. Astr. Soc.*, **340**, 43-51.
Woolfson, M.M. (2004a) *Mon. Not. R. Astr. Soc.*, **354**, 419-426
Woolfson, M.M. (2004b) *Mon. Not. R. Astr. Soc.*, **354**, 1150-1156
Woolfson, M.M., (2007) *Mon. Not. R. Astr. Soc.*, **376**, 1173-1181
Woolfson, M.M. (2011) *On the Origin of Planets*: *By Means of Natural Simple Processes*.(a) pp. 348-350, (b) pp. 269-293. Imperial College Press: London.
Woolfson, M.M. (2013a) *Mon. Not. R. Astr. Soc.*, **436**, 1492
Woolfson, M.M. (2013b) *Earth, Moon and Planets*, **111**, 1-14
Woolfson, M.M. (2016) *Earth, Moon and Planets*, **117**, 77-91
Zel,dovich, Ya. B. and Raiser, Yu. P. (1966) *Physics of Shock Waves and High-Temperature Hydrodynamic Phenomena*. Academic Press: New York